\documentclass[twocolumn]{aastex62}

\newcommand{\Bol}{\mathrm{bol}}
\newcommand{\Clu}{\mathrm{clump}}

\newcommand{\Equ}{\mathrm{Equ}}

\newcommand{\Hyd}{\mathrm{H}}
\newcommand{\Los}{\mathrm{LOS}}
\shorttitle{X-Ray Spectral Model from Clumpy Torus and Its Application to Circinus Galaxy}
\shortauthors{Tanimoto et al.}
\usepackage{amsmath}

\begin{document}
\title{XCLUMPY: X-Ray Spectral Model from Clumpy Torus and Its Application to Circinus Galaxy}
\author{Atsushi Tanimoto}\affiliation{Department of Astronomy, Kyoto University, Kyoto 606-8502, Japan}
\author{Yoshihiro Ueda}\affiliation{Department of Astronomy, Kyoto University, Kyoto 606-8502, Japan}
\author{Hirokazu Odaka}\affiliation{Department of Physics, Faculty of Science, The University of Tokyo, 7-3-1 Hongo, Bunkyo-ku, Tokyo 113-0033, Japan}\affiliation{Kavli IPMU (WPI), UTIAS, The University of Tokyo, 5-1-5 Kashiwanoha, Kashiwa, Chiba 277-8583, Japan}
\author{Toshihiro Kawaguchi}\affiliation{Department of Economics, Management and Information Science, Onomichi City University, Hiroshima 722-8506, Japan}
\author{Yasushi Fukazawa}\affiliation{Department of Physics Science, Hiroshima University, Hiroshima 739-8526, Japan}
\author{Taiki Kawamuro}\affiliation{National Astronomical Observatory of Japan, Tokyo 181-8588, Japan}
\begin{abstract}
We construct an X-ray spectral model from the clumpy torus in an active galactic nucleus (AGN), designated as ``XCLUMPY'', utilizing the Monte Carlo simulation for Astrophysics and Cosmology framework (MONACO: \citealt{Odaka11, Odaka16}). The adopted geometry of the torus is the same as that in \cite{Nenkova08a, Nenkova08b}, who assume a power law distribution of clumps in the radial direction and a normal distribution in the elevation direction. We investigate the dependence of the X-ray continuum and Fe K$\alpha$ fluorescence line profile on the torus parameters. Our model is compared with other torus models: MYTorus model \citep{Murphy09}, Ikeda model \citep{Ikeda09}, and CTorus model \citep{Liu14}. As an example, we also present the results applied to the broadband X-ray spectra of the Circinus galaxy observed with \textit{XMM-Newton}, \textit{Suzaku}, and \textit{NuSTAR}. Our model can well reproduce the data, yielding a hydrogen column density along the equatorial plane $N_{\mathrm{H}}^{\mathrm{Equ}} = 9.08_{-0.08}^{+0.14} \times 10^{24}$ cm$^{-2}$, a torus angular width $\sigma = 14.7_{-0.39}^{+0.44}$ degree, and a 2--10 keV luminosity $\log L_{2-10}/\mathrm{erg \ s^{-1}} = 42.8$. These results are discussed in comparison with the observations in other wavelengths.
\end{abstract}

\section{Introduction}
It is widely accepted that in an active galactic nucleus (AGN) obscuring matter composed of gas and dust surrounds the supermassive black hole (SMBH) \citep[e.g.,][]{Antonucci93, Urry95, Ramos17}. This structure, often referred to as the ``torus'', plays a key role in AGN feeding, serving as a mass reservoir linking the SMBH and host galaxy. It is also suggested that the torus may be produced by outflow from the accretion disk \citep{Elitzur06}, which is an important process of AGN feedback \citep{Fabian12}. Thus, elucidating the torus structure, and ultimately its physical origins, is essential to understand the mechanisms of the SMBH and galaxy co-evolution \citep{Kormendy13}. Nevertheless, many of basic properties of AGN tori still remain unclear.

X-ray observations are a powerful tool to investigate the nature of surrounding material around the SMBH. Unlike radio lines and far to mid-infrared continuum emission, which are sensitive only to cool gas and dust, respectively, X-rays can trace all matter including gas and dust with various physical conditions. The X-ray spectrum of an AGN mainly consists of the direct power law component from the center (emitted via Comptonization of softer photons by a hot corona) and its reflection components from the accretion disk and torus. This torus reflection component carries important information on the structure of the torus. For instance, the relative intensity to the direct component and shape of the reflected continuum, as well as those of the fluorescence Fe K$\alpha$ line at 6.4 keV, strongly depend on the covering factor (solid angle) and total column density of the torus. Also, photoelectric absorption features of the reflection component constrain the overall torus geometry and inclination angle.

Various X-ray spectral models have been used to compare with the observation data. Many authors often employ the \textsf{pexrav} model (continuum only: \citealt{Magdziarz95}), or the \textsf{pexmon} model (continuum plus fluorescence lines: \citealt{Nandra07}), which analytically calculate the reflection component from cold gas in a semi-infinite (i.e., optically thick) plane. These analytic models are convenient to use but give only rough approximation of the real data, because the geometry of a torus is more complex than a single plane and it is not trivial to assume a Compton-thick reflector.

To take account of more realistic torus geometry, numerical spectral models based on Monte Carlo ray-tracing simulations have been developed (MYTorus model: \citealt{Murphy09}. Ikeda model: \citealt{Ikeda09}. Torus model: \citealt{Brightman11a}). \cite{Murphy09} adopt a bagel-like (i.e., toroidal) shape whose opening angle is fixed. \cite{Ikeda09} and \cite{Brightman11a} assume essentially spherical geometry with two bipolar conical holes, where the opening angle of the torus can be varied as a free parameter (see \citealt{Liu15, Balokovic18} for a note and an update of the original model by \citealt{Brightman11a}, respectively). All these models assume a uniform density of the gas, and hence we refer to them as ``smooth torus models". These models have also been widely used to fit the observed X-ray spectra of AGNs (Ikeda model: e.g., \citealt{Awaki09, Eguchi11, Tazaki11, Tazaki13, Kawamuro13, Kawamuro16a, Kawamuro16b, Tanimoto16, Tanimoto18, Oda17, Oda18, Yamada18}, MYTorus model: e.g., \citealt{Yaqoob12, Yaqoob15, Yaqoob16, Koss13, Koss15, Koss16, Koss17, Ricci14a, Ricci14b, Ricci15, Ricci16a, Ricci16b, Ricci17a, Ricci17b}, Torus model: e.g., \citealt{Brightman11b, Brightman12, Brightman13, Brightman14, Brightman15, Brightman16, Brightman17, Buchner14, Buchner15, Gandhi14, Gandhi15, Gandhi17}).

Many observations indicate, however, that AGN tori must be composed of dusty clumps rather than of smooth gas \citep[e.g.,][]{Krolik88, Wada02, Honig07}. Here we summarize 4 major pieces of observational evidence for clumpy tori: (1) geometrical thickness of tori, (2) 10 $\mu$m silicate emission feature in the infrared spectra of Seyfert 2 galaxies, (3) correlations between the infrared and X-ray luminosities, and (4) X-ray spectral feature of torus-reflection components:

\begin{enumerate}
\item The velocity dispersion of matter must be as large as a typical rotation velocity of the torus ($V \sim 100$ km s$^{-1}$) in order to sustain a geometrically thick torus as inferred from the absorbed AGN fraction \citep[e.g.,][]{Ricci15}. In the case of smooth gas, it corresponds to the thermal velocity. However, the velocity of dusty gas cannot become so fast. This is because dust grains would reach a maximum sublimation temperature ($T \sim 1500$ K: \citealt{Laor93}).
\item \cite{Pier92, Pier93} constructed infrared spectral models from a smooth torus and predicted that the 10 $\mu$m silicate feature would appear only in absorption in Seyfert 2 galaxies. By contrast, \cite{Mason09, Nikutta09} detected silicate emission line features in the infrared spectra of Seyfert 2 galaxies.
\item An inner region of the torus has a higher temperature since dust is heated by radiation from the central accretion disk. In smooth tori, it is therefore expected that the ratio between the infrared and X-ray luminosities would be systematically higher in type 1 AGNs where inner parts of the torus are more visible. In reality, however, the correlations are very similar among type 1 and type 2 AGNs \citep{Gandhi09, Ichikawa12, Ichikawa17, Ichikawa19, Asmus15, Kawamuro16a}.
\item The X-ray spectra of heavily absorbed AGNs often show unabsorbed torus-reflection components \citep[e.g.][]{Ueda07}. If we apply smooth torus models to such X-ray spectrum, the inclination angle and torus half-opening angle become very close to each other \citep{Awaki09, Eguchi11, Tazaki11, Tanimoto16, Tanimoto18}. This corresponds to the geometry where the observer sees the AGN through the edge of the torus boundary and seems unrealistic in a statistical sense. Clumpy tori can naturally explain these features because it predict that a significant fraction of the unabsorbed reflection component from the far-side torus.
\end{enumerate}

In the infrared band, \cite{Nenkova08a, Nenkova08b} constructed spectral models from clumpy tori, by assuming a power law distribution in the radial direction and a normal distribution in the elevation direction for the configuration of clumps. This model, called CLUMPY, has been successfully applied to the infrared spectra of nearby AGNs (see \citealt{Ramos17} and references therein). \cite{Stalevski12, Stalevski16} also computed the infrared spectra from a clumpy torus of a two-phase medium with slightly different geometry from that in the CLUMPY model.

It is only recently that X-ray spectral models from clumpy torus have been developed (CTorus model: \citealt{Liu14}. Furui model: \citealt{Furui16}). \cite{Liu14} constructed such a model, using the Geant4 library \citep{Agostinelli03, Allison06, Allison16} for the first time. They adopted clump distribution confined in a partial sphere. Later, \cite{Furui16} made an X-ray clumpy torus model with bagel-like geometry by using the Monte Carlo simulation for astrophysics and cosmology framework (MONACO: \citealt{Odaka11, Odaka16}), which also utilizes the Geant4 library but is optimized for astrophysical applications. In this model, Compton down-scattering of fluorescence lines, which produces ``Compton shoulder" in the line profile, is properly taken into account.

In this paper, we construct a new X-ray clumpy torus model designated as ``XCLUMPY", by adopting the same geometry of clump distribution as that of the CLUMPY model in the infrared band \citep{Nenkova08a, Nenkova08b}. This enables us to directly compare the results inferred from the infrared and X-ray bands, which constrain the spatial distribution of dust and that of all matter including gas, respectively. The structure of this paper is as follows. Section 2 and 3 describe the adopted torus geometry and the details of Monte Carlo simulations, respectively. In Section 4, we present major results of our model such as dependencies of the X-ray continuum and Fe K$\alpha$ line profile on the torus parameters. We also compare our model with other torus models: MYTorus model \citep{Murphy09}, Ikeda model \citep{Ikeda09}, and CTorus model \citep{Liu14}. In Section 5, we apply our model to the broadband X-ray spectra of the Circinus galaxy observed with \textit{XMM-Newton}, \textit{Suzaku},\textit{NuSTAR}. Throughout the paper we adopt the cosmological parameters ($H_{0} = 70$ km s$^{-1}$ Mpc$^{-1}$, $\Omega_{\mathrm{m}} = 0.3$, $\Omega_{\lambda} = 0.7$). The errors on the spectral parameters correspond to the 90\% confidence limits for a single parameter.

\begin{deluxetable*}{llll}
\tablecaption{Summary of Parameters}
\tablehead
{
Note                                                            & Parameter					                            	&
Grid						                                    & Units}
\startdata
(01)					                                        & $r_{\mathrm{in}}$ 		                            	&
$0.05$				                                    	    & pc		                            					\\
(02)					                                	    & $r_{\mathrm{out}}$	                            		&
$1.00$				                                       	    & pc				                                		\\
(03)					                                	    & $R_{\mathrm{clump}}$		                                &
$0.002$				                                    	    & pc					                            		\\
(04)					                                   	    & $N_{\mathrm{clump}}^{\Equ}$                           	&
$10.0$				                                    	    & \nodata				                            		\\
(05)					                                	    & $q$			                            				&
$0.50$				                                    	    & \nodata				                            		\\
(06)						                                    & $\sigma$					                            	&
$10.0,20.0,30.0,40.0,50.0,60.0,70.0$    	                    & degree			                            			\\
(07)					                                   	    & $\log N_{\Hyd}^{\Equ}/$cm$^{-2}$	                     	&
$22.00,22.25,22.50,22.75,23.00,23.25,23.50,23.75$               &       		                                        	\\
                                                                &                                                           &
$24.00,24.25,24.50,24.75,25.00,25.25,25.50,25.75,26.00$         & \nodata                                                   \\
(08)				                                		    & $i$					                            		&
$18.2,31.8,41.4,49.5,56.6,63.3,69.5,75.5,81.4,87.1$             & degree				                            		\\
(09)					                                	    & $\Gamma$			                            			&
$1.50,1.60,1.70,1.80,1.90,2.00,2.10,2.20,2.30,2.40,2.50$        & \nodata					                            	\\
(10)					                                	    & $\log E_{\mathrm{cut}}/$keV	                            &
$1.00,1.50,2.00,2.50,3.00$		                                &
\enddata
\tablecomments{Column (01): inner radius of the torus. Column (02): outer radius of the torus. Column (03): radius of the clump. Column (04): number of the clump along the equatorial plane. Column (05): index of the radial density profile. Column (06): torus angular width. Column (07): hydrogen column density along the equatorial plane. Column (08): inclination angle. Column (09): photon index. Column (10): cutoff energy.}
\end{deluxetable*}
\begin{figure}
\plotone{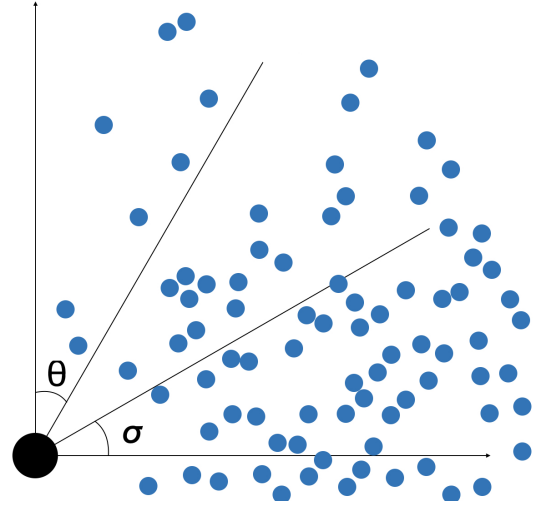}
\caption{Cross section view of the torus. We assume a power law distribution of clumps in the radial direction and a normal distribution in the elevation direction.}
\end{figure}

\section{Torus Geometry}
In our model, a torus is not continuous medium, but is composed of many clumps randomly distributed following a given number density function. For simplicity, each clump is a sphere with a radius of $R_{\mathrm{clump}}$, and has a uniform hydrogen number density $n_{\mathrm{H}}$. As mentioned in Section 1, we adopt the same geometry as that in \cite{Nenkova08a, Nenkova08b}, who assumed a power law distribution of clumps in the radial direction between inner and outer radii, and a normal distribution in the elevation direction (Figure 1). Specifically, the number density function $d(r,\theta,\phi)$ (in units of pc$^{-3}$) is represented in the spherical coordinate system (where $r$ is radius, $\theta$ is polar angle, and $\phi$ is azimuth) as:
\begin{equation}
d(r,\theta,\phi) = N \left(\frac{r}{r_{\mathrm{in}}}\right)^{-q} \exp{\left(-\frac{(\theta-\pi/2)^2}{\sigma^2}\right)}.
\end{equation}
where $N$ is the normalization, $q$ is the index of the radial density profile, and $\sigma$ is the torus angular width.

The normalization $N$ is related to the number of clumps along the equatorial plane $N_{\mathrm{clump}}^{\mathrm{Equ}}$ as
\begin{align}
N_{\mathrm{clump}}^{\mathrm{Equ}}   & = \int_{r_{\mathrm{in}}}^{r_{\mathrm{out}}} d\left(r,\frac{\pi}{2},0\right)\pi R_{\mathrm{clump}}^2 dr.\nonumber\\
N                                   & = \frac{(1-q)N_{\mathrm{clump}}^{\mathrm{Equ}}}{\pi R_{\mathrm{clump}}^2 r_{\mathrm{in}}^{q}(r_{\mathrm{out}}^{1-q}-r_{\mathrm{in}}^{1-q})}
\end{align}

By substituting Equation (2) into Equation (1), we obtain the number of clumps along the line-of-sight at a given polar angle: 
\begin{equation}
N_{\mathrm{clump}}^{\mathrm{LOS}}(\theta) = N_{\mathrm{clump}}^{\mathrm{Equ}} \exp{\left(-\frac{(\theta-\pi/2)^2}{\sigma^2}\right)}.
\end{equation}
For instance, for $N_{\mathrm{clump}}^{\mathrm{Equ}} = 10$ and $\sigma = 60$ degree, $N_{\mathrm{clump}}^{\mathrm{LOS}}(\pi/2) \approx 10$, $N_{\mathrm{clump}}^{\mathrm{LOS}}(\pi/3) \approx 7$, and $N_{\mathrm{clump}}^{\mathrm{LOS}}(\pi/6) \approx 3$.

The total number of clumps in the torus $N_{\mathrm{clump}}^{\mathrm{Tot}}$ is obtained by integrating the number density function,
\begin{align}
N_{\mathrm{clump}}^{\mathrm{Tot}}   & = \int_{r_{\mathrm{in}}}^{r_{\mathrm{out}}} \int_{0}^{\pi}\int_{0}^{2\pi} d(r,\theta,\phi) r^2 \sin{\theta} dr d\theta d\phi\nonumber\\
                                    & = \frac{2N_{\mathrm{clump}}^{\mathrm{Equ}}(1-q)(r_{\mathrm{out}}^{3-q}-r_{\mathrm{in}}^{3-q})}{R_{\mathrm{clump}}^2(3-q) (r_{\mathrm{out}}^{1-q}-r_{\mathrm{in}}^{1-q})}\nonumber\\
                                    & \times \int_{0}^{\pi}\exp{\left(-\frac{(\theta-\pi/2)^2}{\sigma^2}\right)}\sin{\theta}d\theta
\end{align}
Typically $N_{\mathrm{clump}}^{\mathrm{Tot}} \sim 10^6$ for the adopted parameters (Table 1).

We also define the total hydrogen column density along the equatorial plane:
\begin{equation}
N_{\mathrm{H}}^{\mathrm{Equ}} = \frac{4}{3}R_{\mathrm{clump}} N_{\mathrm{clump}}^{\mathrm{Equ}}n_{\mathrm{H}}
\end{equation}
Here $\frac{4}{3}R_{\mathrm{clump}} (= \frac{4}{3}\pi R_{\mathrm{clump}}^3/(\pi R_{\mathrm{clump}}^2))$ corresponds to the average length crossing the line-of-sight in one clump (a sphere with a radius of $R_{\mathrm{clump}}$) when the clumps are randomly located. The parameter $N_{\mathrm{H}}^{\mathrm{Equ}}$ can be directly compared with the total optical-depth along the equatorial plane introduced in \cite{Nenkova08a, Nenkova08b}.

To summarize, our model has 8 independent parameters that define the torus properties: (1) inner radius of the torus ($r_{\mathrm{in}}$), (2) outer radius of the torus ($r_{\mathrm{out}}$), (3) radius of each clump ($R_{\mathrm{clump}}$), (4) number of clumps along the equatorial plane ($N_{\mathrm{clump}}^{\mathrm{Equ}}$), (5) index of the radial density profile ($q$), (6) torus angular width ($\sigma$), (7) hydrogen column density along the equatorial plane ($N_{\mathrm{H}}^{\mathrm{Equ}}$), and (8) the inclination angle ($i$, the polar angle of the line of sight).

In our work, we fix (1) $r_{\mathrm{in}}$, (2) $r_{\mathrm{out}}$, (4) $N_{\mathrm{clump}}^{\mathrm{Equ}}$, and (5) $q$ to the mean values obtained by \cite{Ichikawa15}, who applied the CLUMPY model to the infrared spectral energy distribution of nearby 21 AGNs. We note that the clump-size parameter (3) $R_{\mathrm{clump}}$ does not affect the calculation of the infrared spectra as long as it is sufficiently small \citep{Nenkova08a, Nenkova08b}. This is not the case for the X-ray spectra, however. We therefore assume a typical value (a logarithmic average) within the torus region based on a theoretical estimate by \cite{Kawaguchi10, Kawaguchi11}. The adopted value (0.002 pc) is compatible with the observations of transient X-ray absorption events by torus clumps in nearby AGNs \citep{Markowitz14}. Thus, our model has three free parameters related to the torus: (6) $\sigma$, (7) $N_{\mathrm{H}}^{\mathrm{Equ}}$, and (8) $i$. Table 1 summarizes the values of the fixed parameters and the range of the free parameters. Although $N_{\mathrm{clump}}^{\mathrm{Equ}}$ and $q$ are fixed in our table model, we examine the dependencies of the spectra on these parameters by running simulations for limited sets of parameters in Appendices A and B (see also Section 4.3).

\begin{figure*}
\plottwo{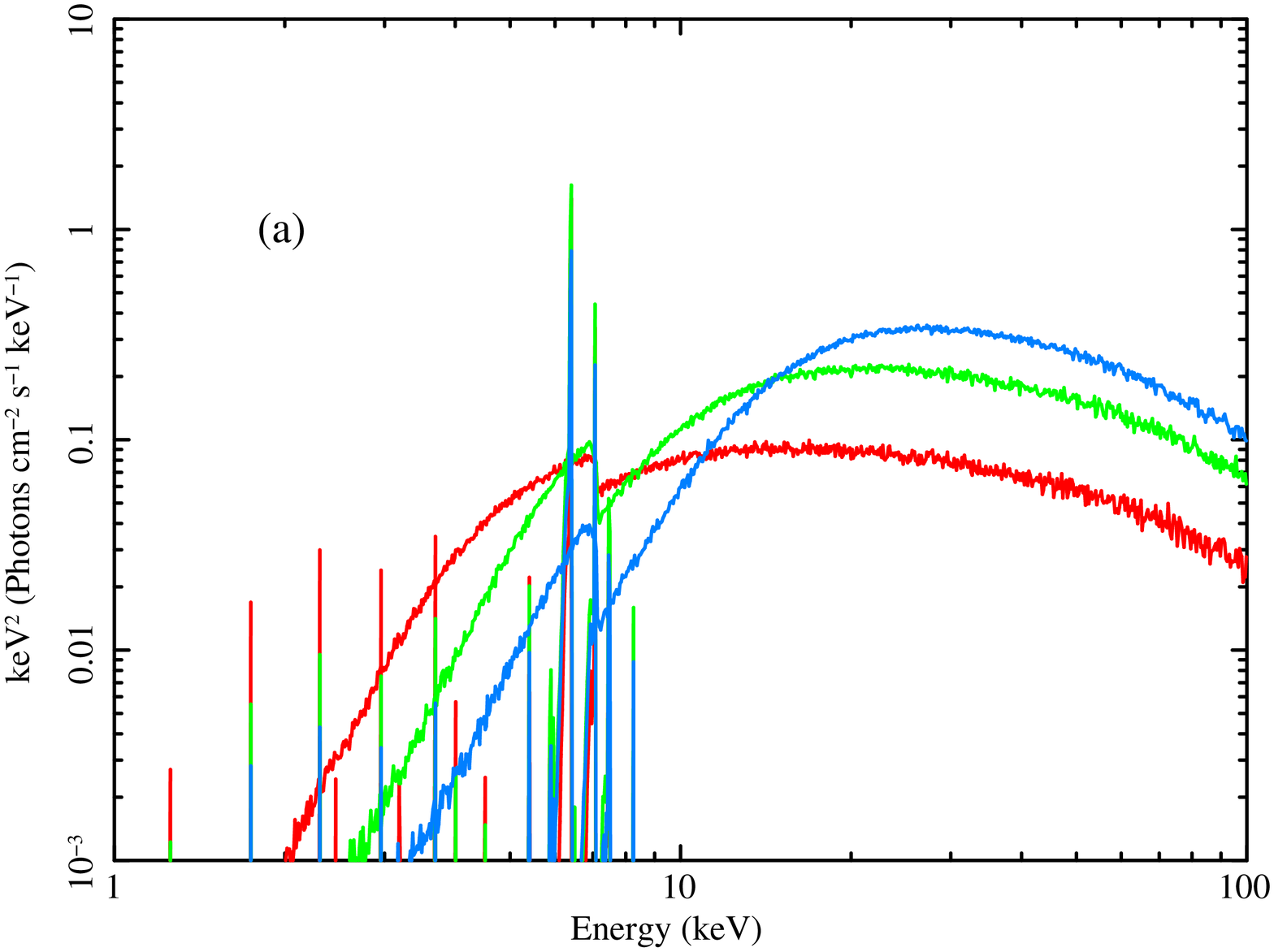}{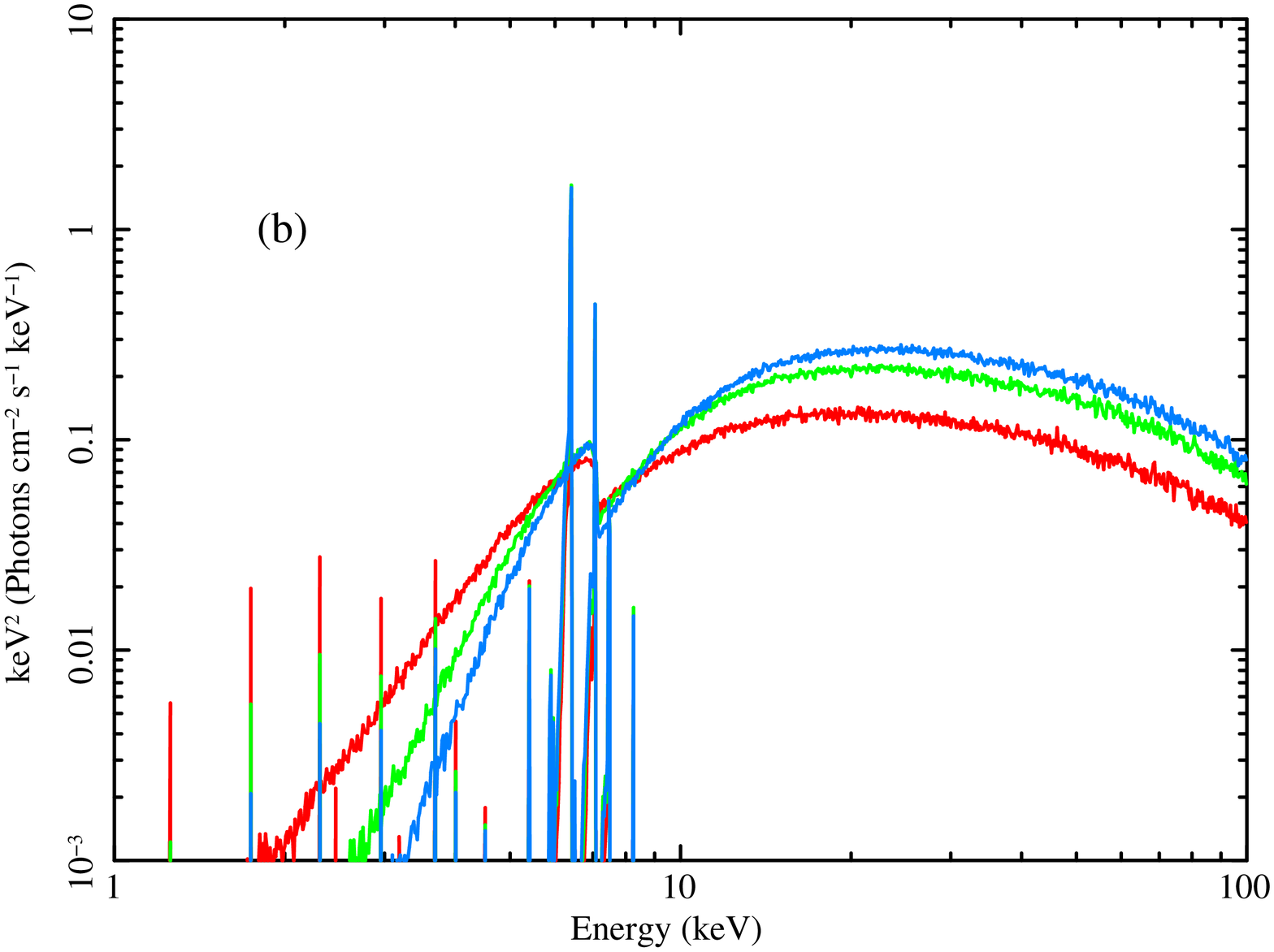}
\epsscale{0.5}
\plotone{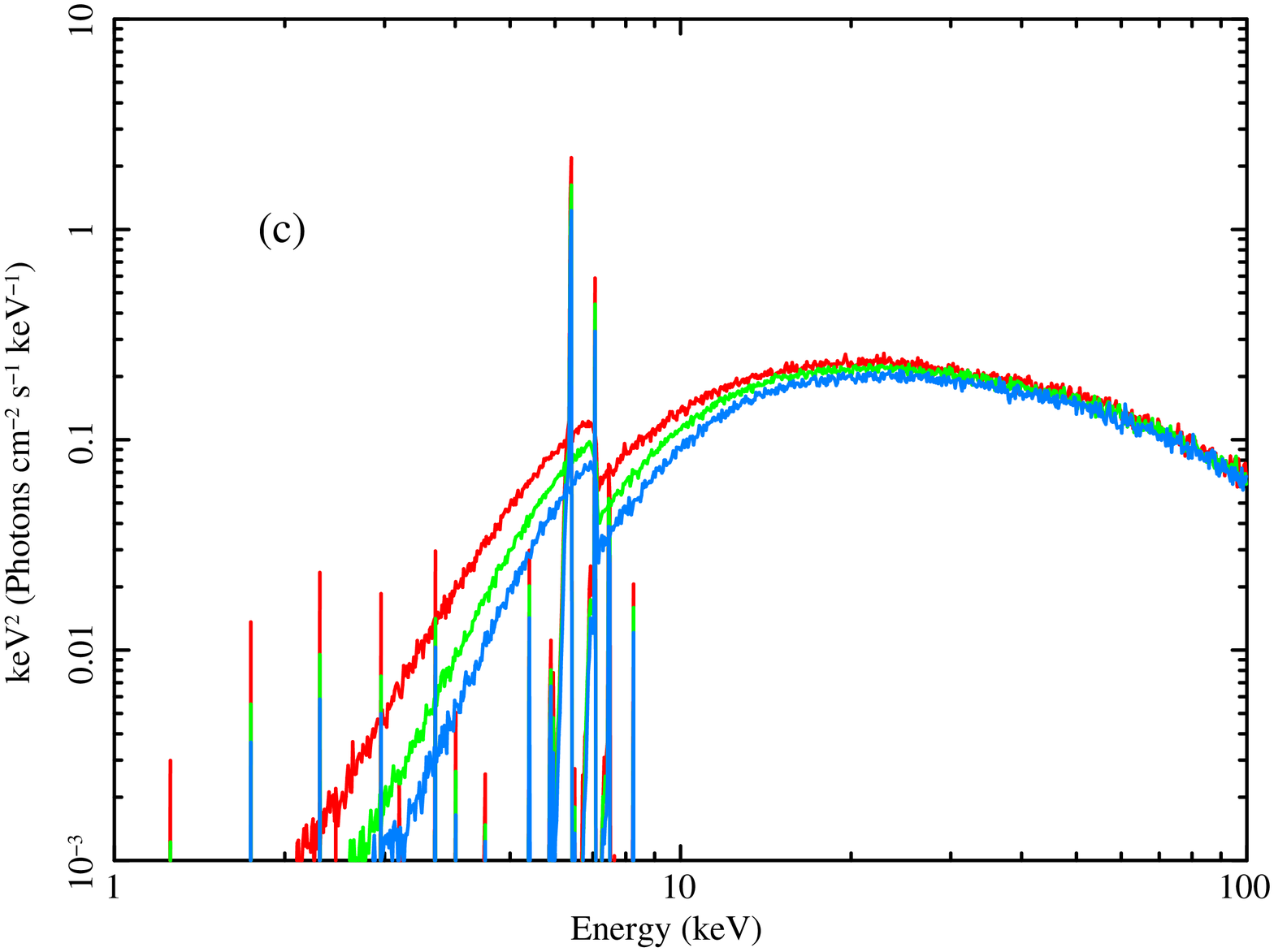}
\caption{Dependence of the reflected X-ray spectrum on the (a) hydrogen column density along the equatorial plane, (b) torus angular width, and (c) inclination angle. We adopt the following values as default parameters: $\log N_{\mathrm{H}}/$cm$^{-2} = 24.0$, $\sigma = 40.0$ degree, $i = 60.0$ degree, $\Gamma = 2.0$, and $E_{\mathrm{cut}} = 100$ keV. (a) Red line: $\log N_{\mathrm{H}}/$cm$^{-2} = 23.5$. Green line: $\log N_{\mathrm{H}}/$cm$^{-2} = 24.0$. Blue line: $\log N_{\mathrm{H}}/$cm$^{-2} = 24.5$. (b) Red line: $\sigma = 20.0$ degree. Green line: $\sigma = 40.0$ degree. Blue line: $\sigma = 60.0$ degree. (c) Red line: $i = 40.0$ degree. Green line: $i = 60.0$ degree. Blue line: $i = 80.0$ degree.}
\end{figure*}
\begin{figure*}
\plottwo{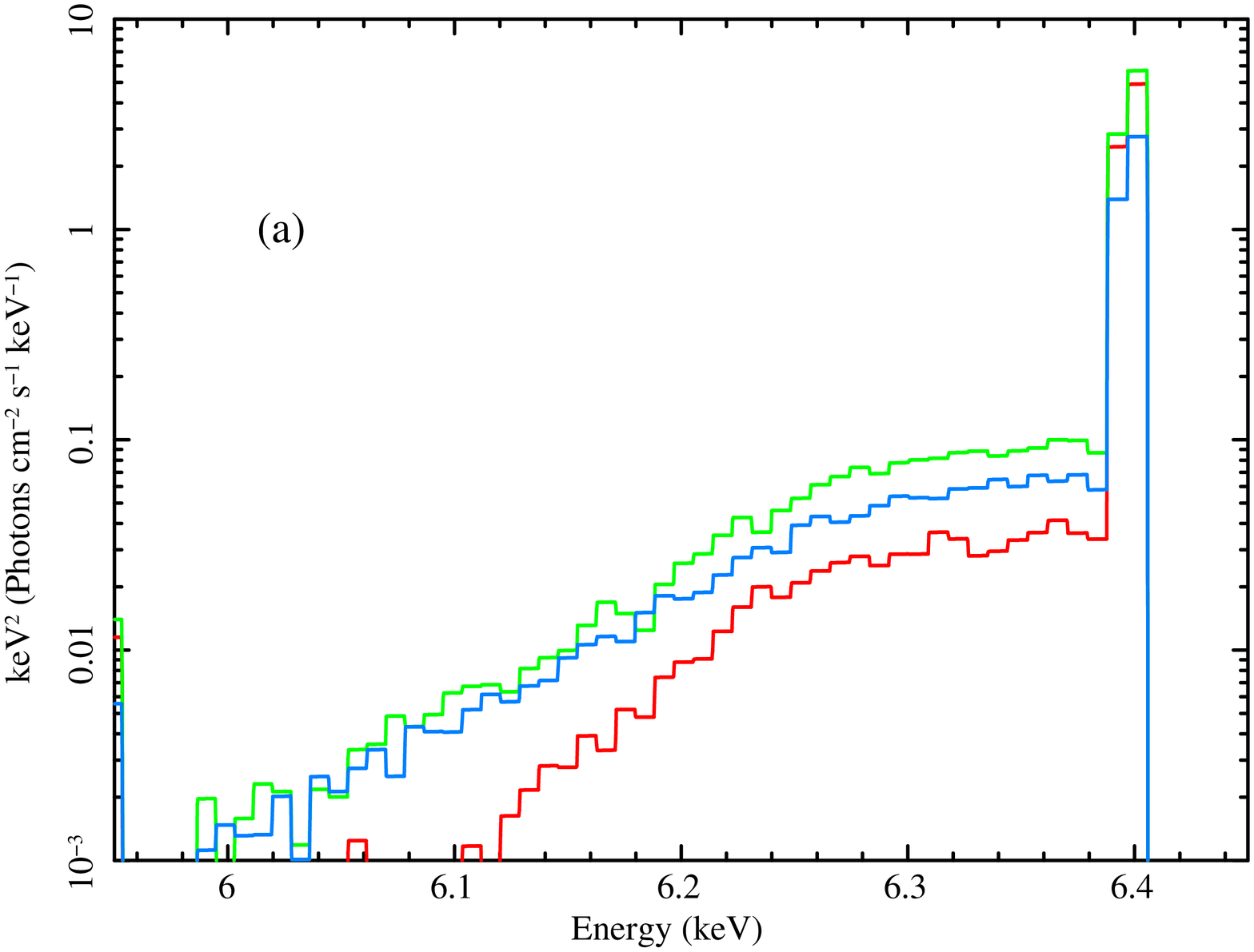}{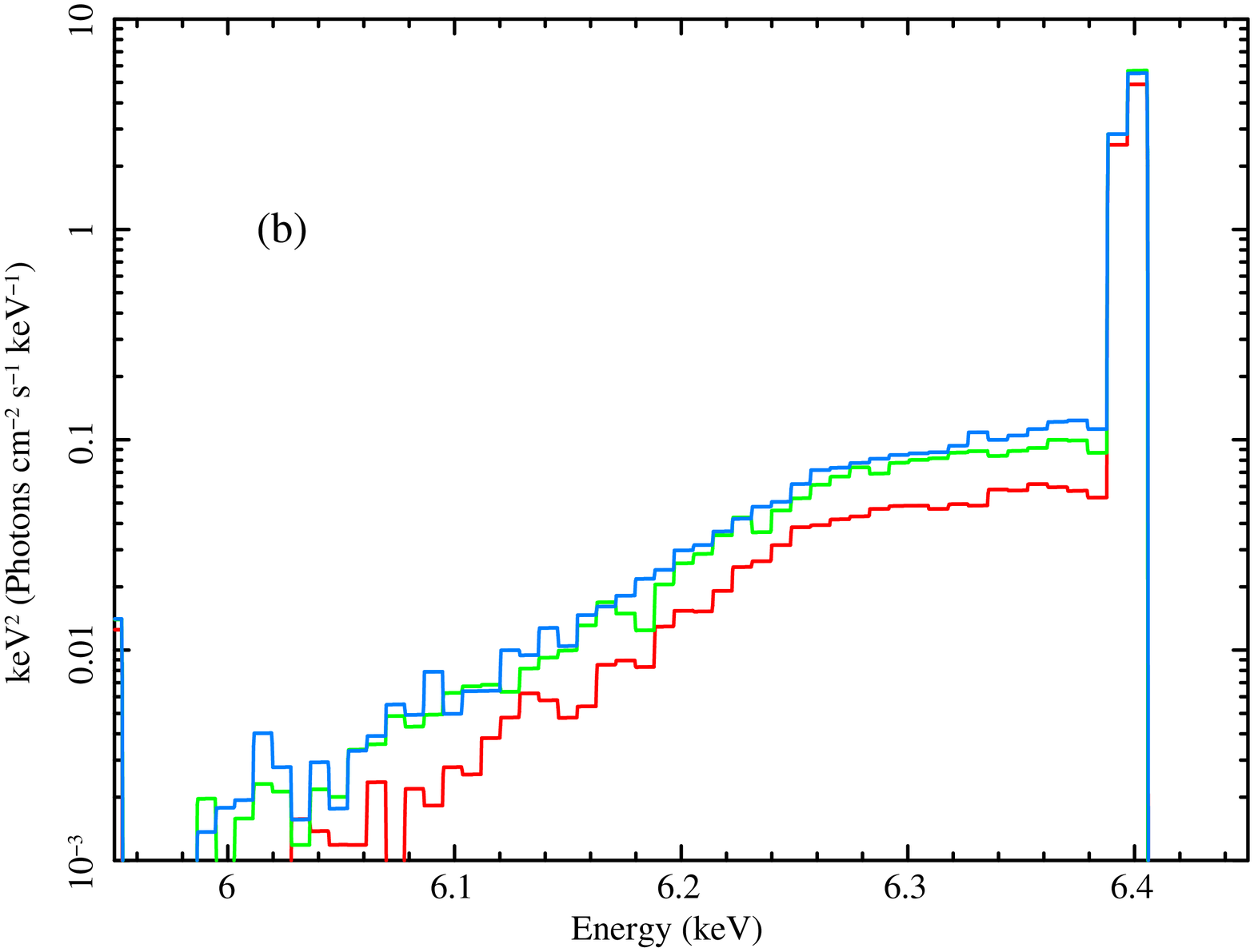}
\epsscale{0.5}
\plotone{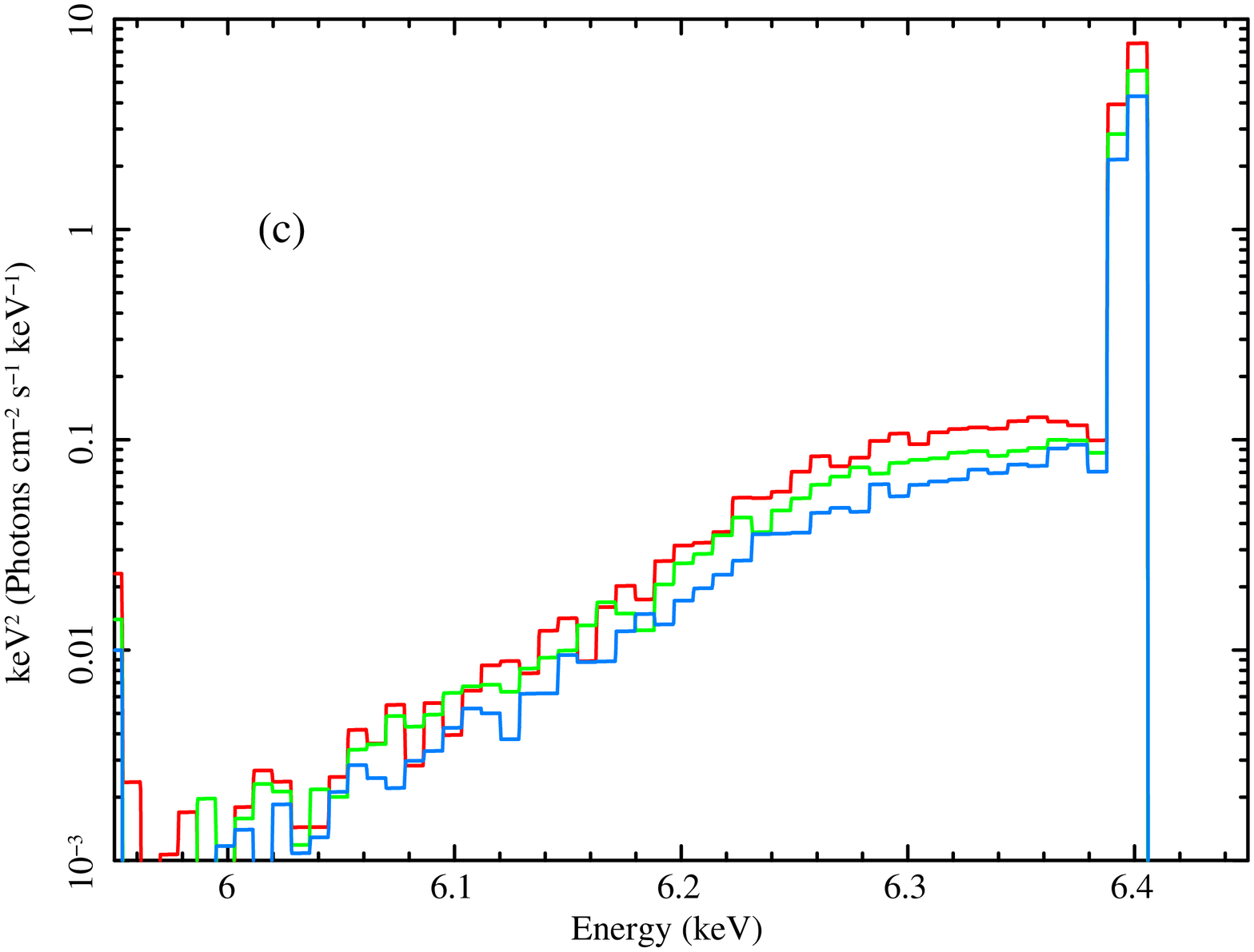}
\caption{Dependence of the Fe K$\alpha$ line profile on the (a) hydrogen column density along the equatorial plane, (b) torus angular width, and (c) inclination angle. We adopt the following values as default parameters: $\log N_{\mathrm{H}}/$cm$^{-2} = 24.0$, $\sigma = 40.0$ degree, $i = 60.0$ degree, $\Gamma = 2.0$, and $E_{\mathrm{cut}} = 100$ keV. (a) Red line: $\log N_{\mathrm{H}}/$cm$^{-2} = 23.5$. Green line: $\log N_{\mathrm{H}}/$cm$^{-2} = 24.0$. Blue line: $\log N_{\mathrm{H}}/$cm$^{-2} = 24.5$. (b) Red line: $\sigma = 20.0$ degree. Green line: $\sigma = 40.0$ degree. Blue line: $\sigma = 60.0$ degree. (c) Red line: $i = 40.0$ degree. Green line: $i = 60.0$ degree. Blue line: $i = 80.0$ degree.}
\end{figure*}
\begin{figure*}
\plottwo{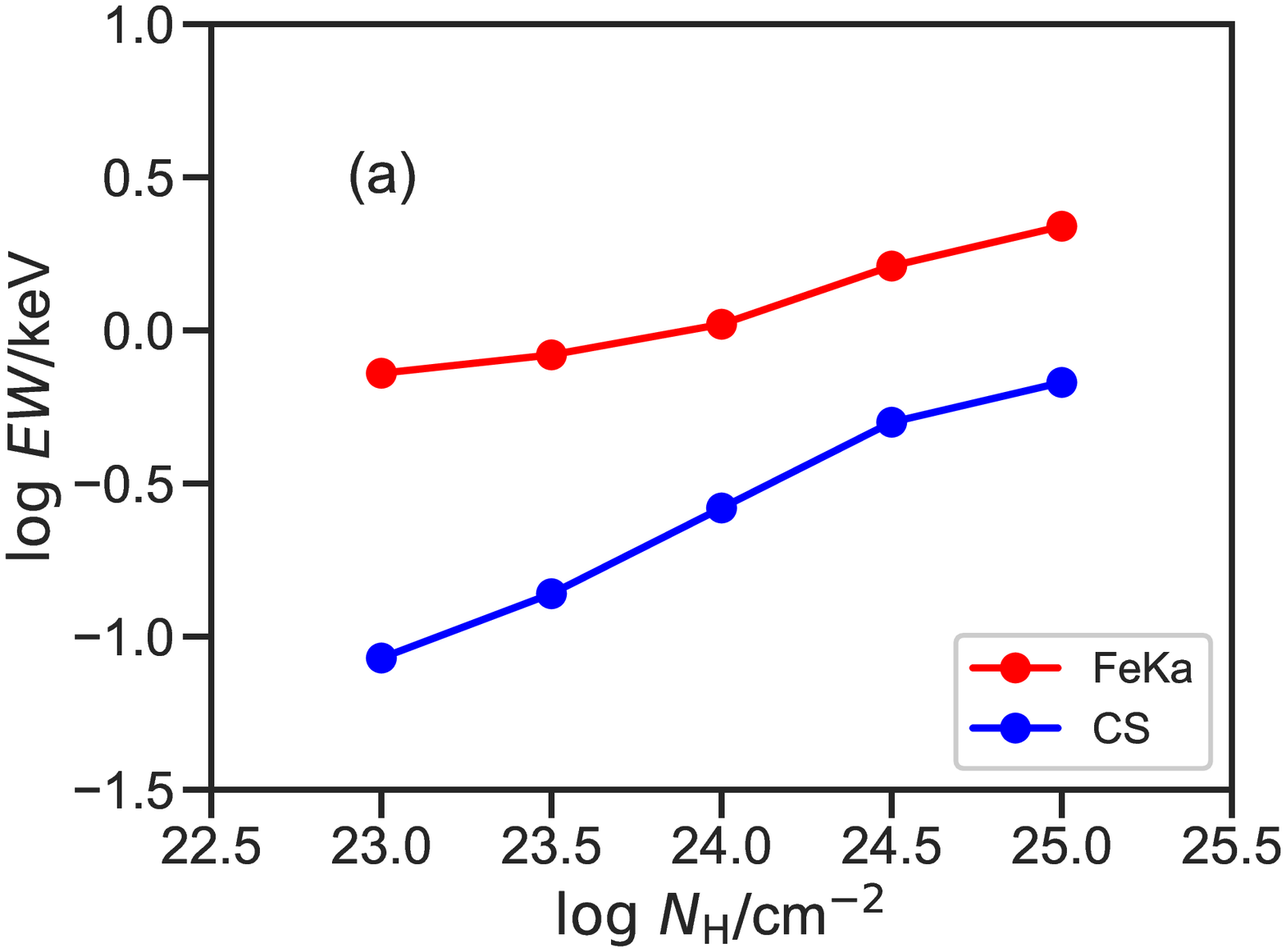}{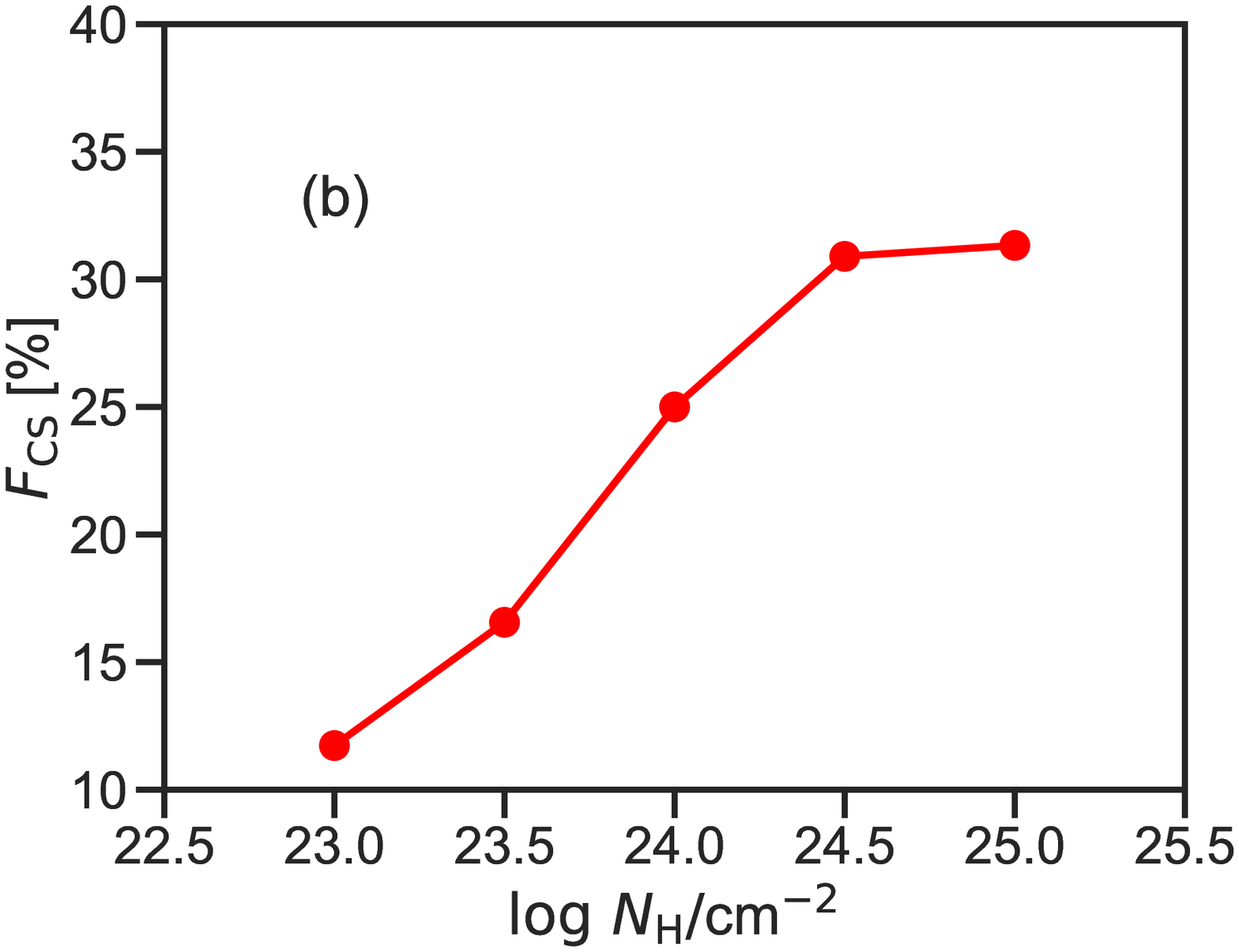}
\plottwo{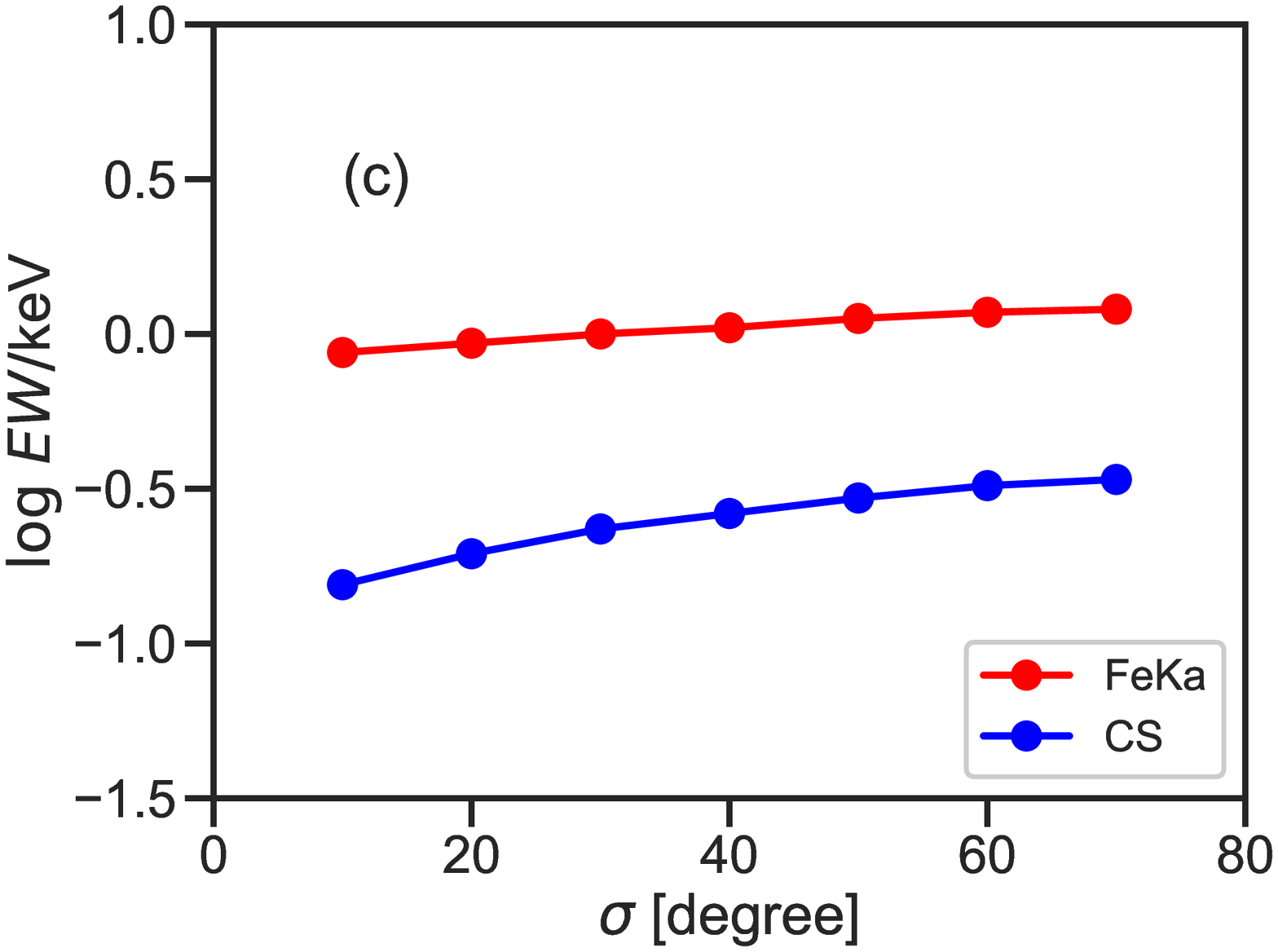}{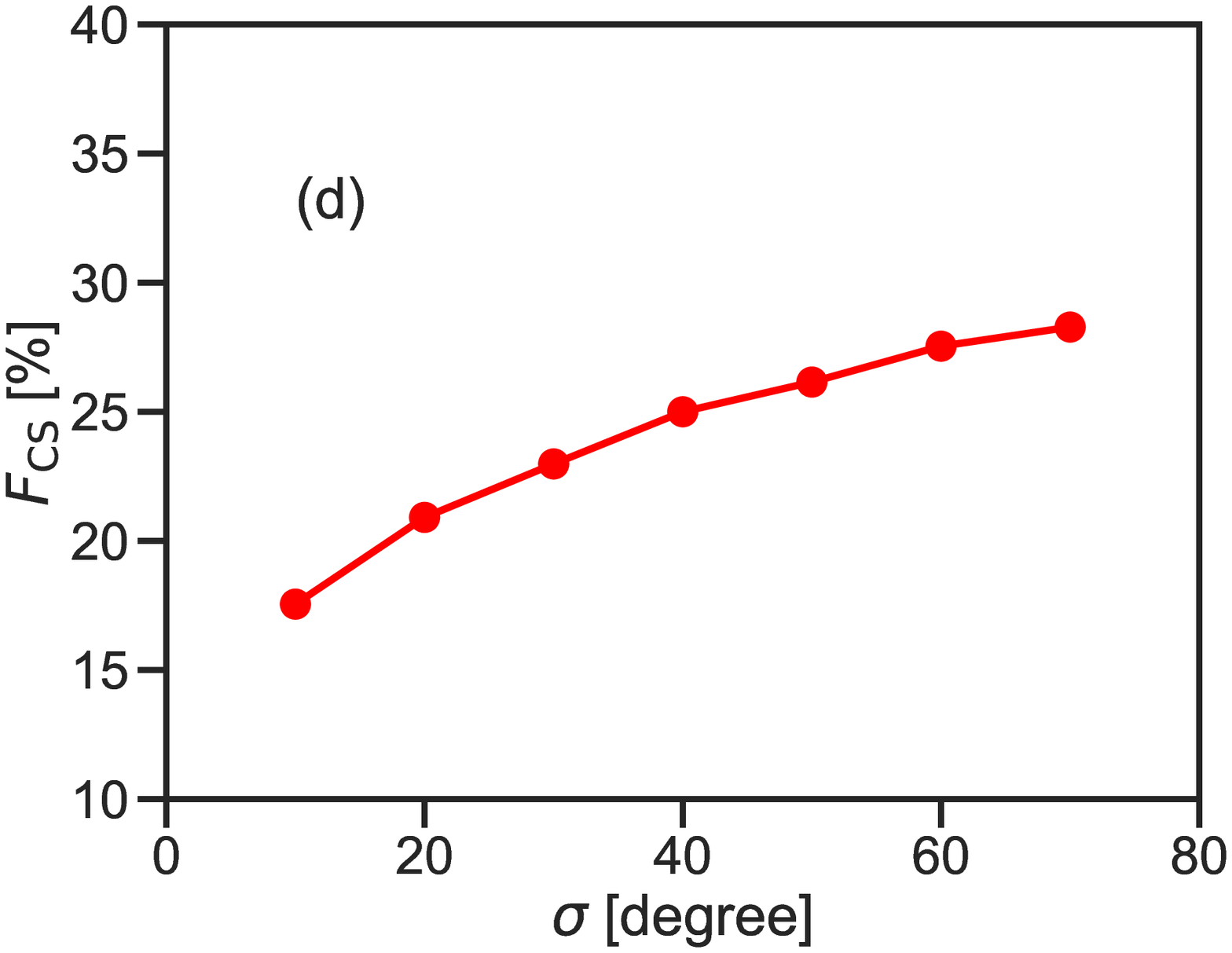}
\plottwo{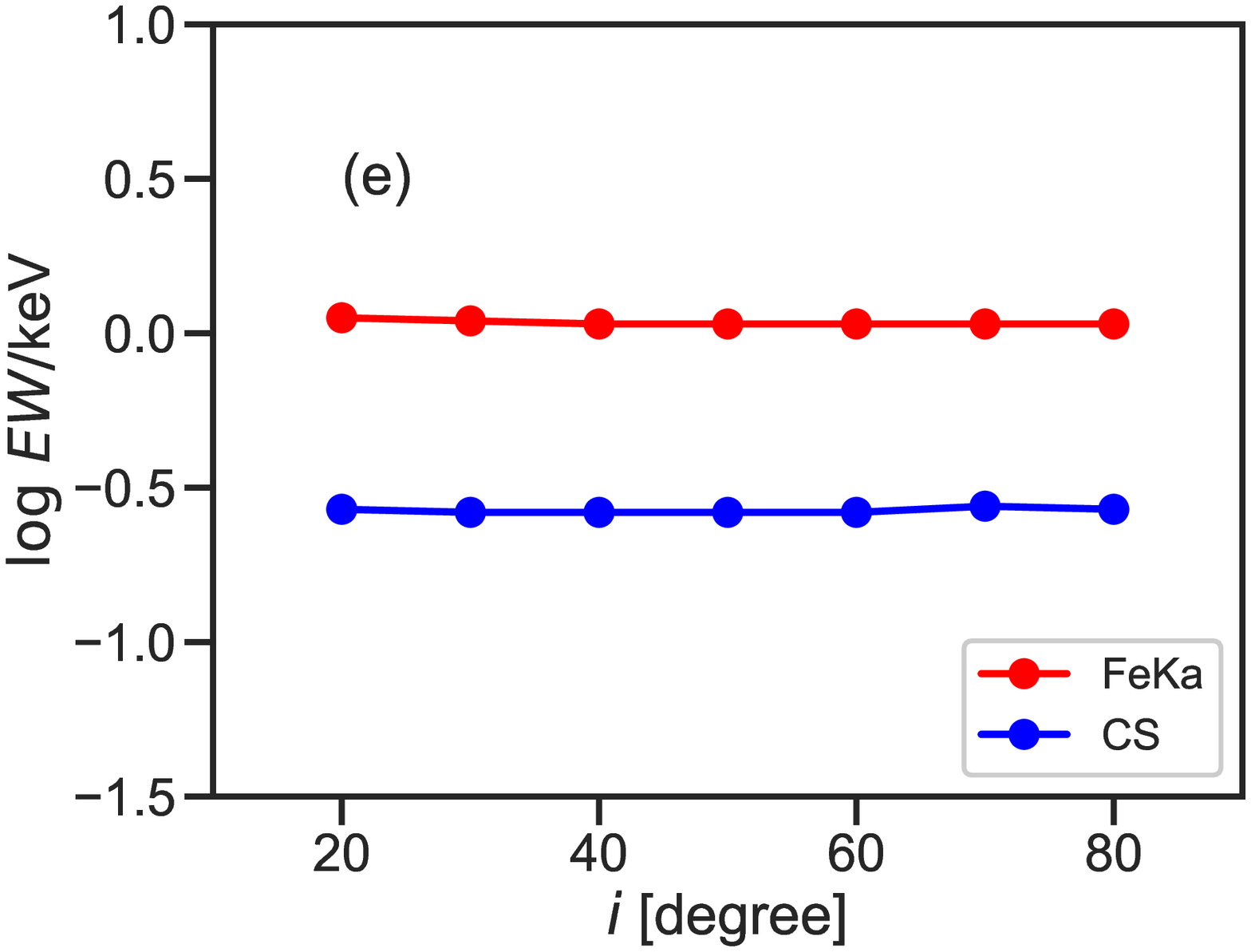}{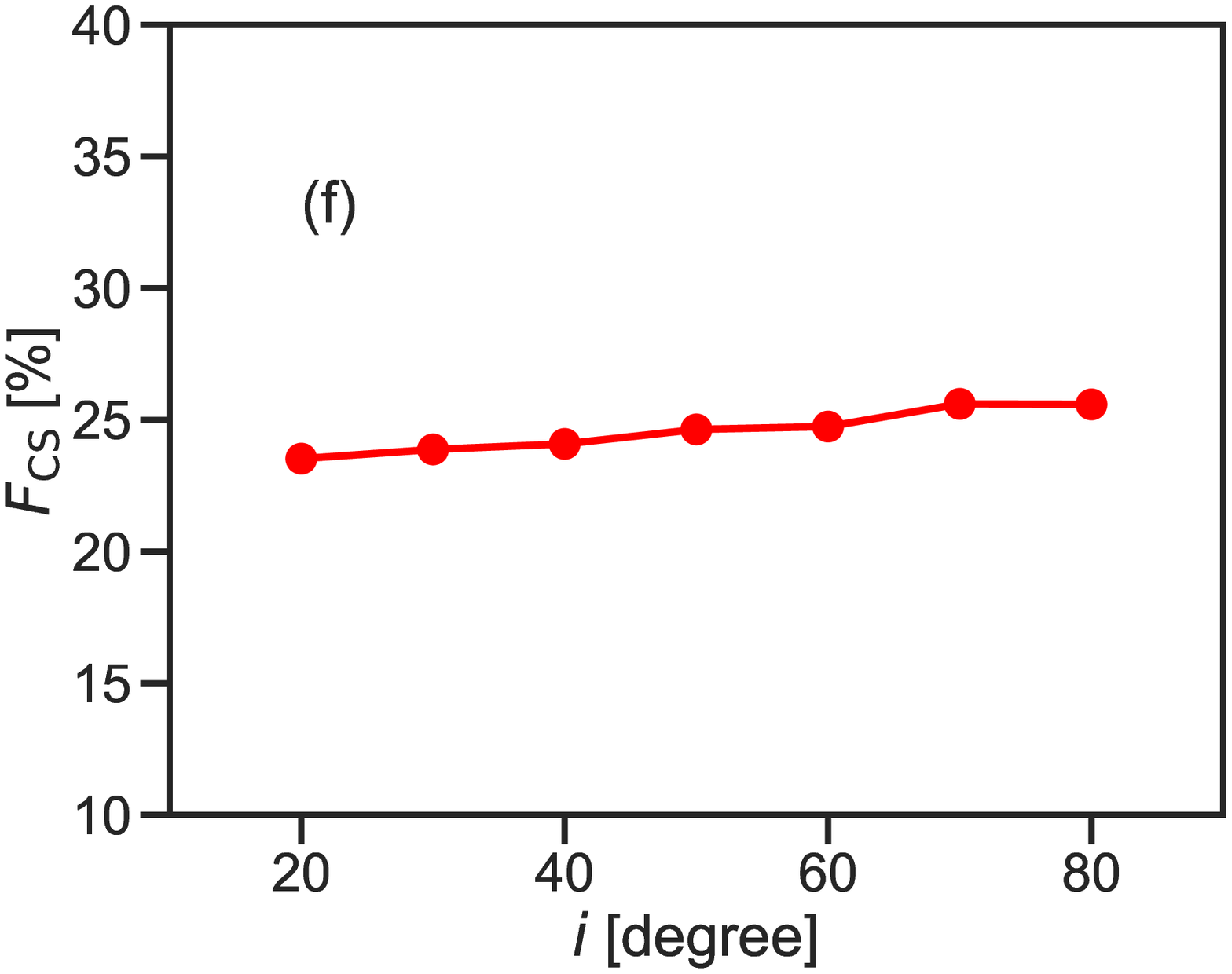}
\caption{Dependence of equivalent widths and Compton shoulder fraction on the (a)-(b) hydrogen column density along the equatorial plane, (c)-(d) torus angular width, and (e)-(f) inclination angle. We set basic torus parameters: $\log N_{\mathrm{H}}/$cm$^{-2} = 24.0$, $\sigma = 40.0$ degree, $i = 60.0$ degree, $\Gamma = 2.0$, and $E_{\mathrm{cut}} = 100$ keV. Left: Red line: equivalent width of Fe K$\alpha$. Blue line: equivalent width of Compton shoulder. Right: Red line: Compton shoulder fraction.}
\end{figure*}
\begin{figure*}
\plottwo{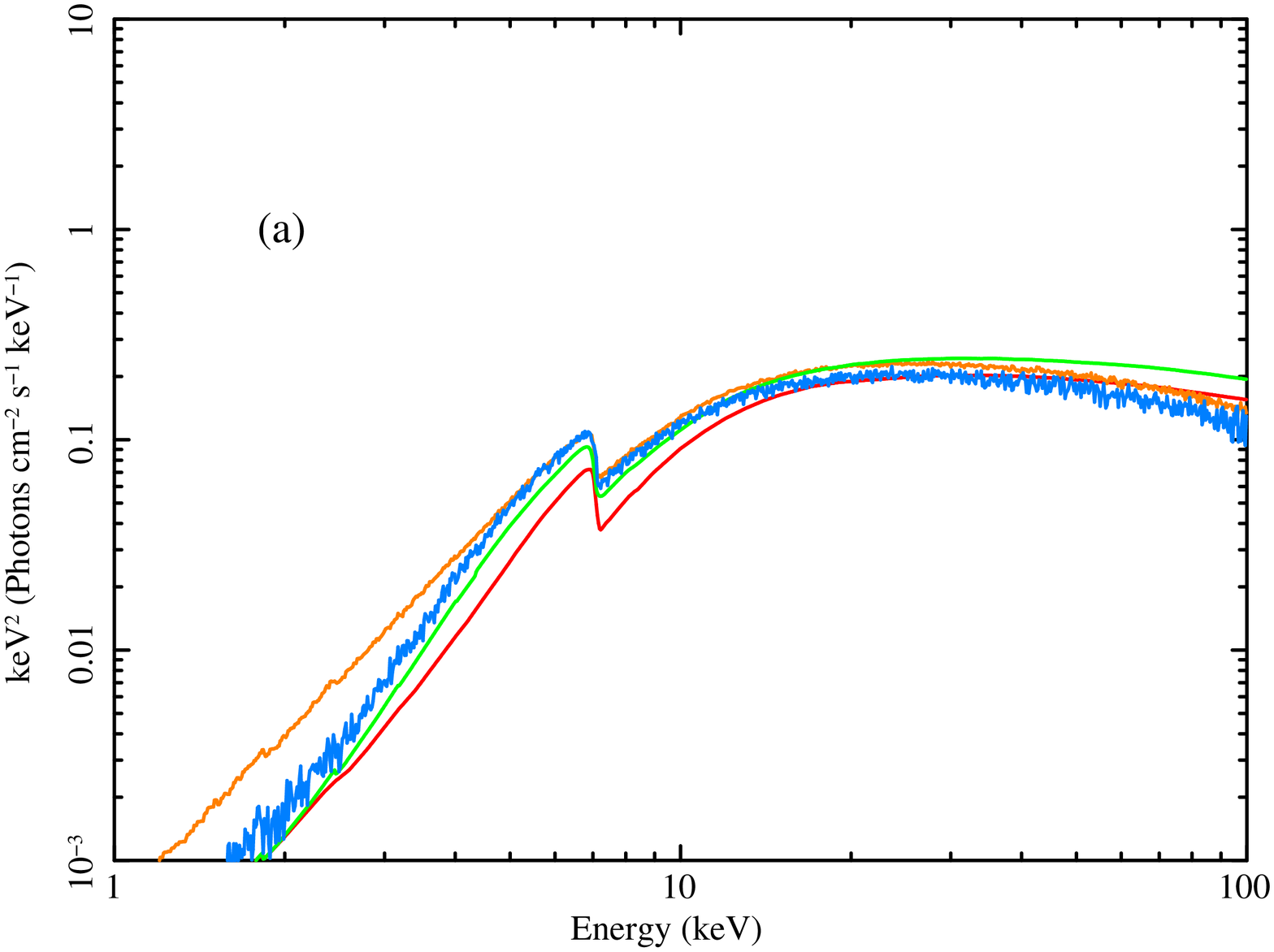}{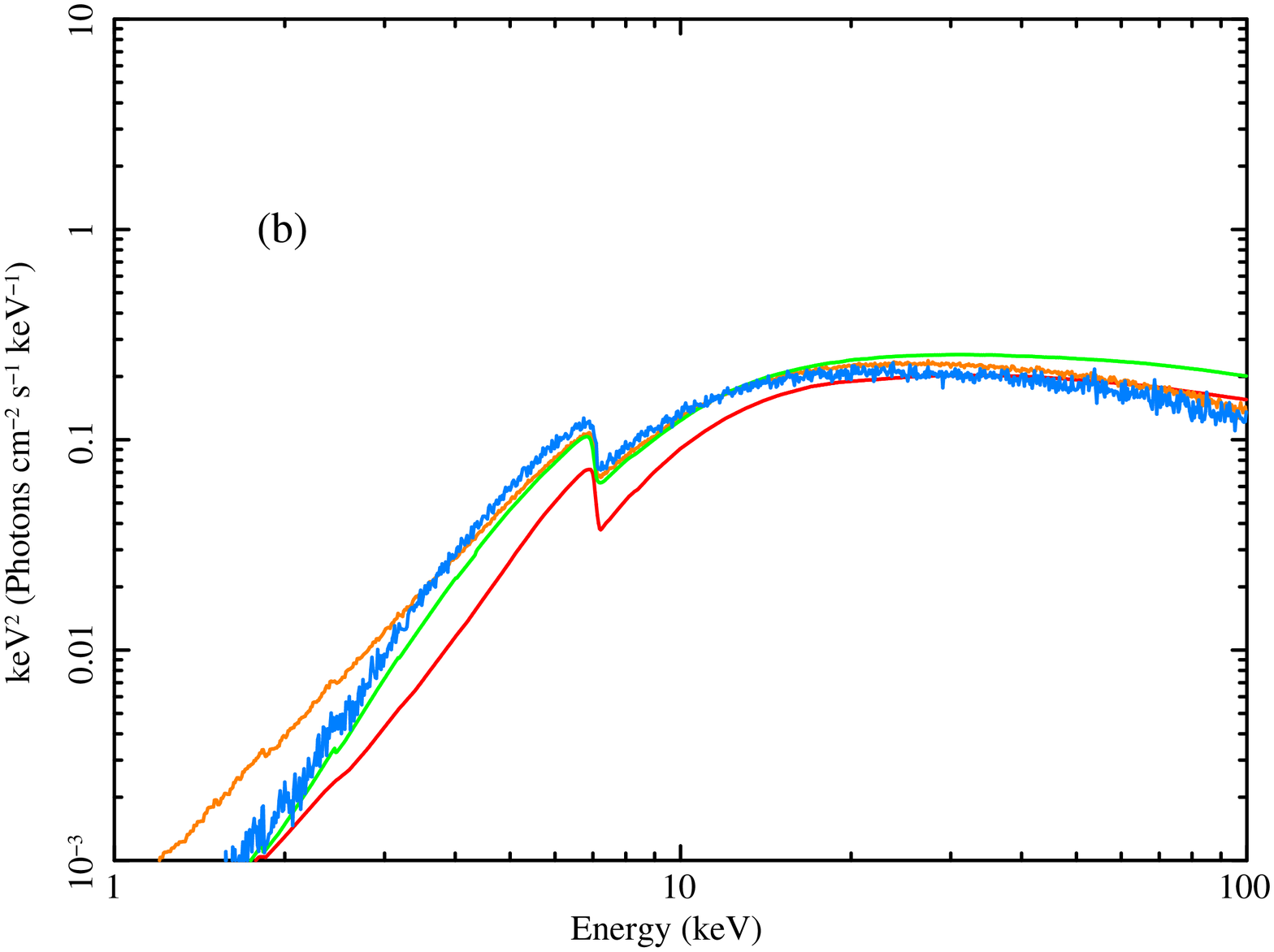}
\plottwo{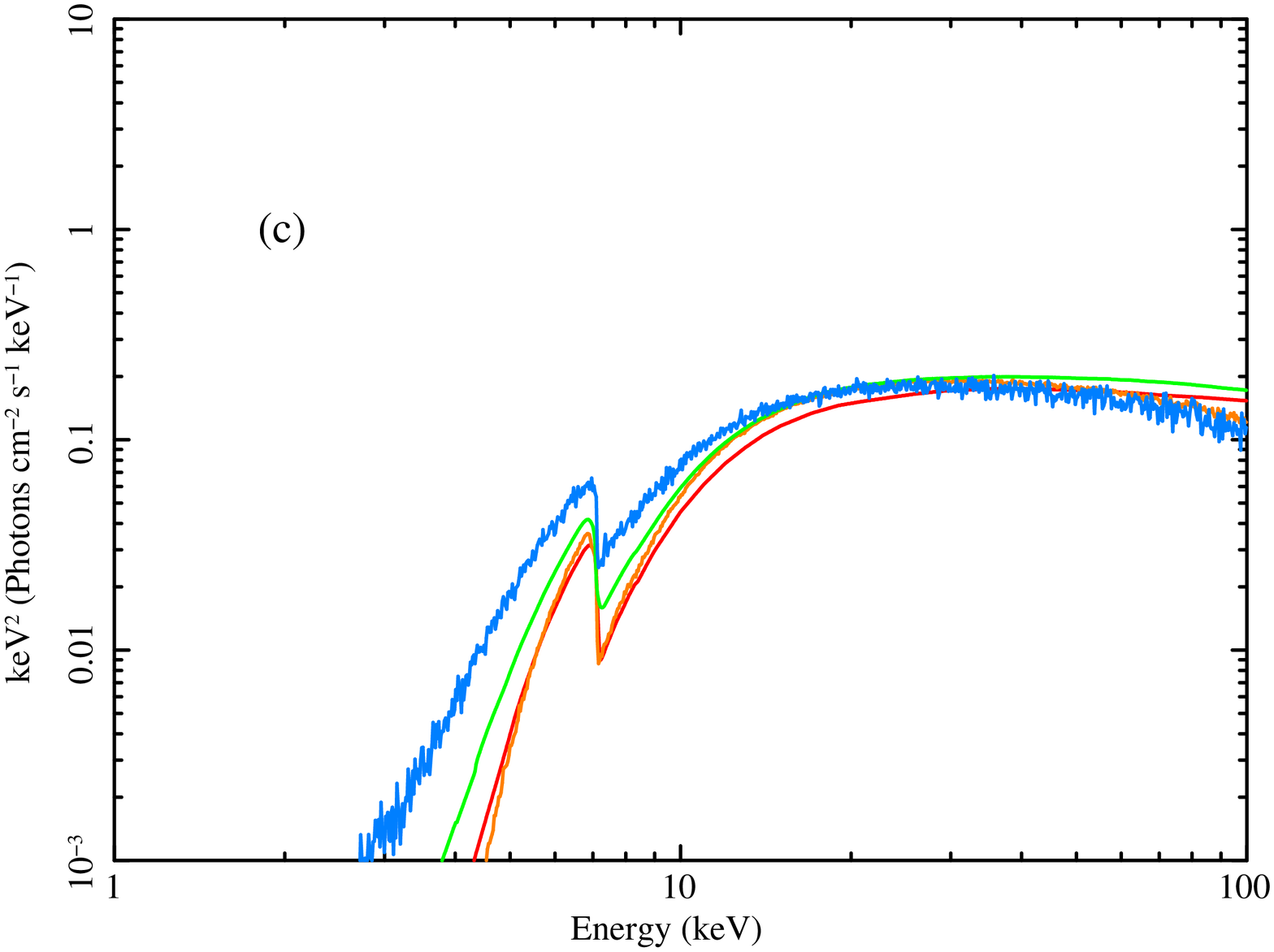}{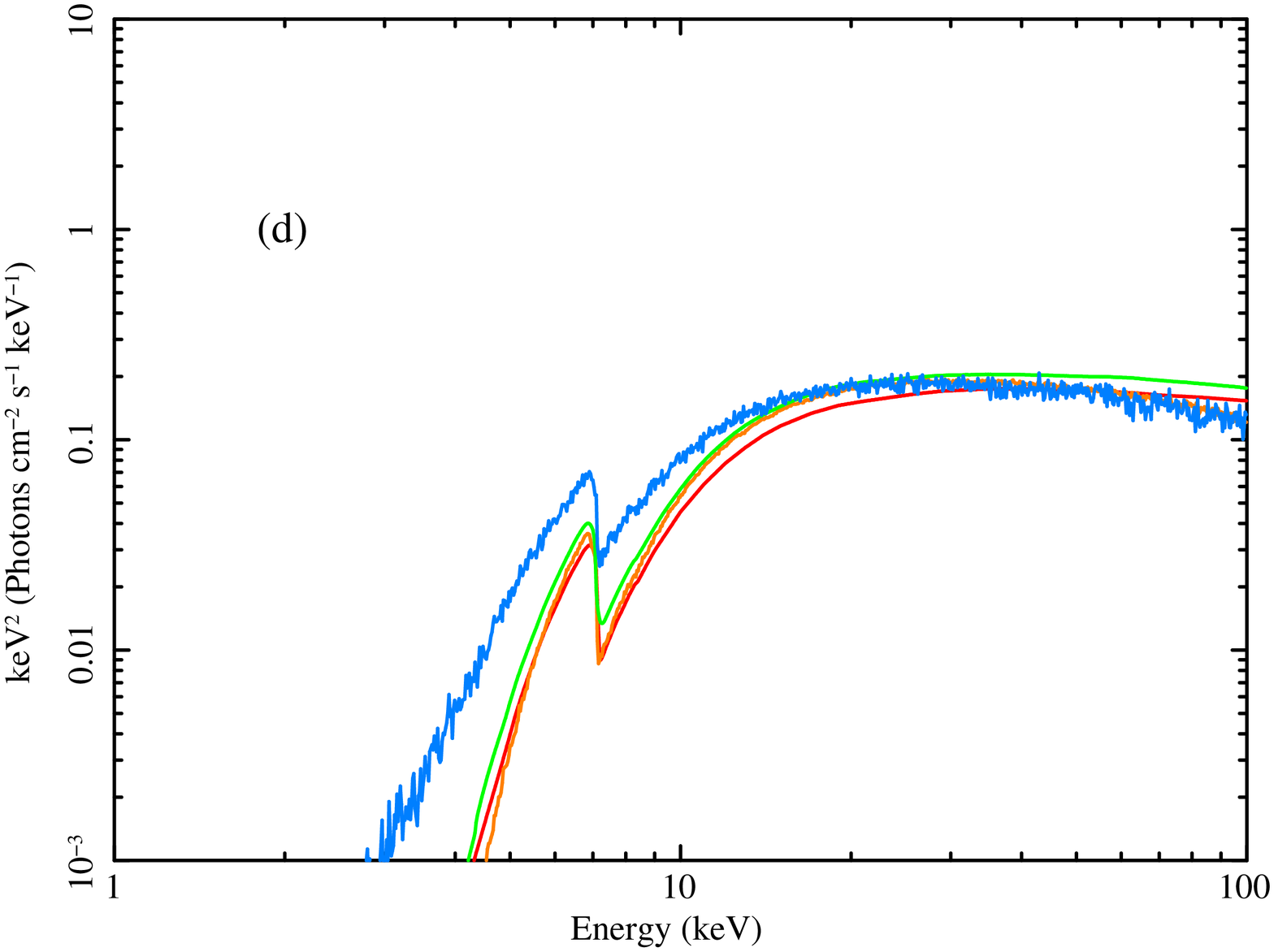}
\caption{Comparison of the reflected X-ray continuum among the torus models. Left: $N_{\mathrm{clump}}^{\mathrm{Equ}} = 5.0$. Right: $N_{\mathrm{clump}}^{\mathrm{Equ}} = 10.0$. Top: $i = 20.0$ degree. Bottom: $i = 87.0$ degree. Red line: MYTorus model. Orange line: Ikeda model. Green line: CTorus model. Blue line: XCLUMPY model. Torus parameters are as follows: MYTorus model: $N_{\mathrm{H}}^{\mathrm{Equ}} = 1.0 \times 10^{24}$ cm$^{-2}$, $\theta_{\mathrm{open}} = 60.0$ degree (fixed). Ikeda model: $N_{\mathrm{H}}^{\mathrm{Equ}} = 1.0 \times 10^{24}$ cm$^{-2}$, $\theta_{\mathrm{open}} = 60.0$ degree. CTorus model: $N_{\mathrm{H}}^{\mathrm{Equ}} = 1.5 \times 10^{24}$ cm$^{-2}$, $\theta_{\mathrm{open}} = 60.0$ degree (fixed). XCLUMPY model: $N_{\mathrm{H}}^{\mathrm{Equ}} = 1.0 \times 10^{24}$ cm$^{-2}$, $\sigma = 30.0$ degree.}
\end{figure*}
\begin{figure*}
\plottwo{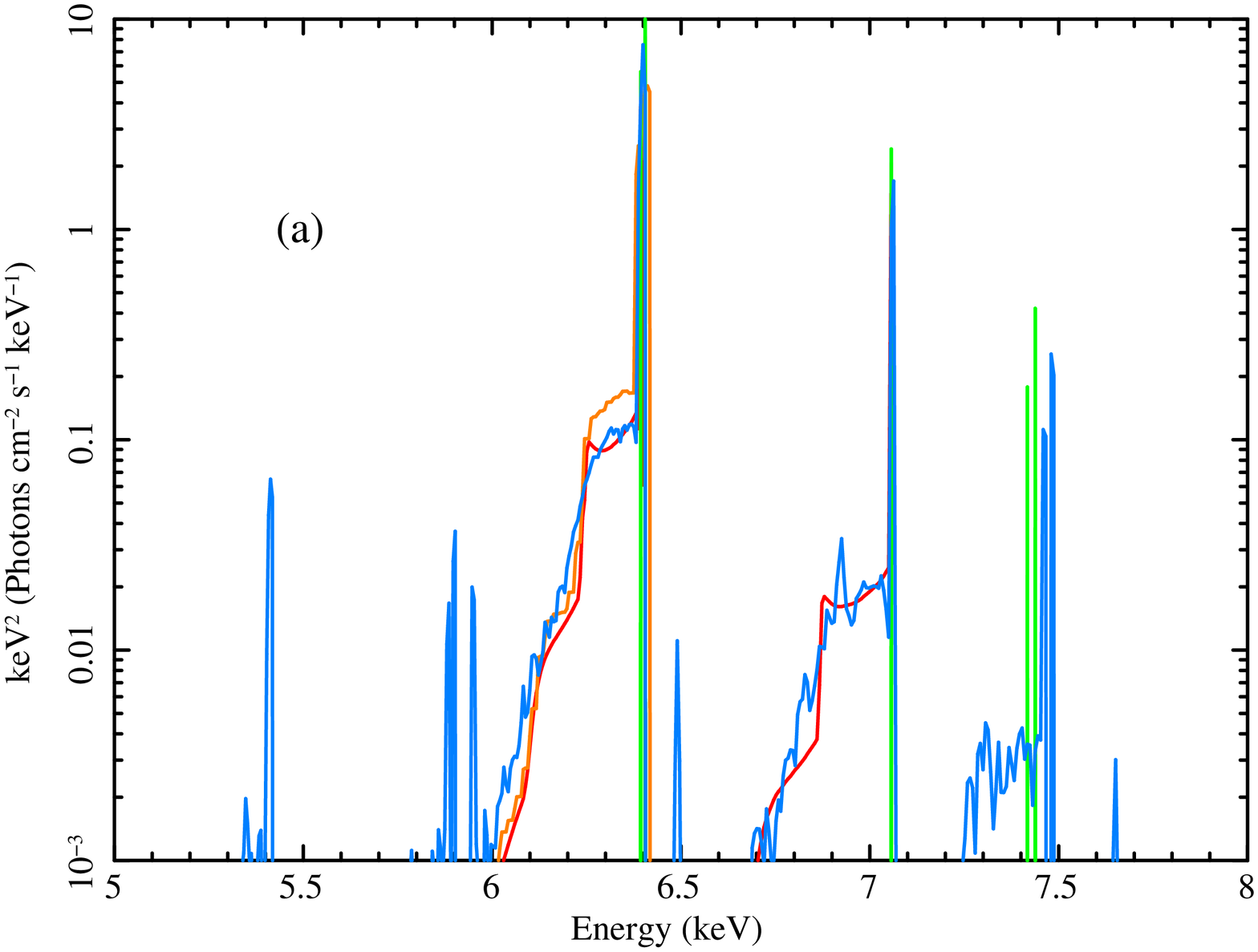}{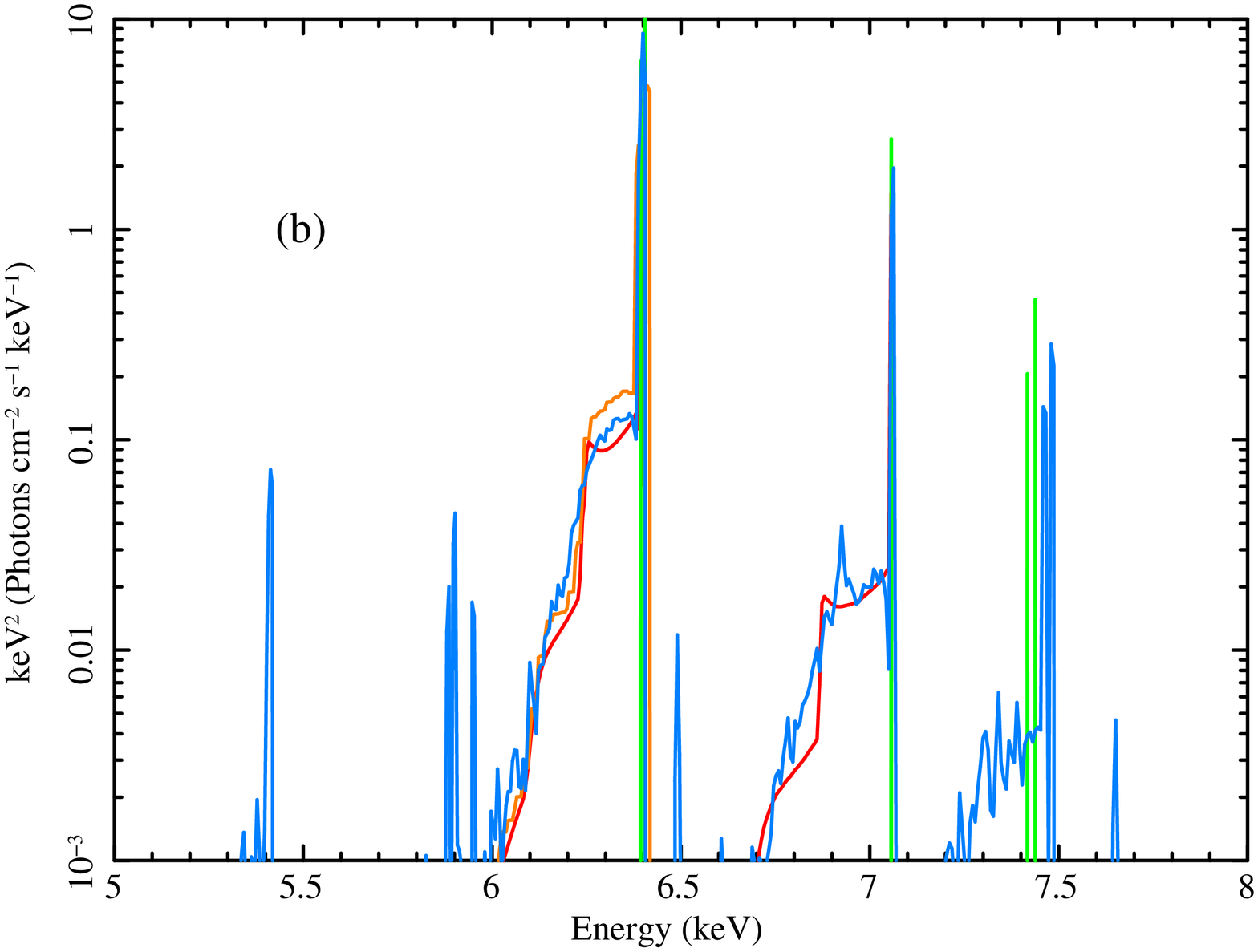}
\plottwo{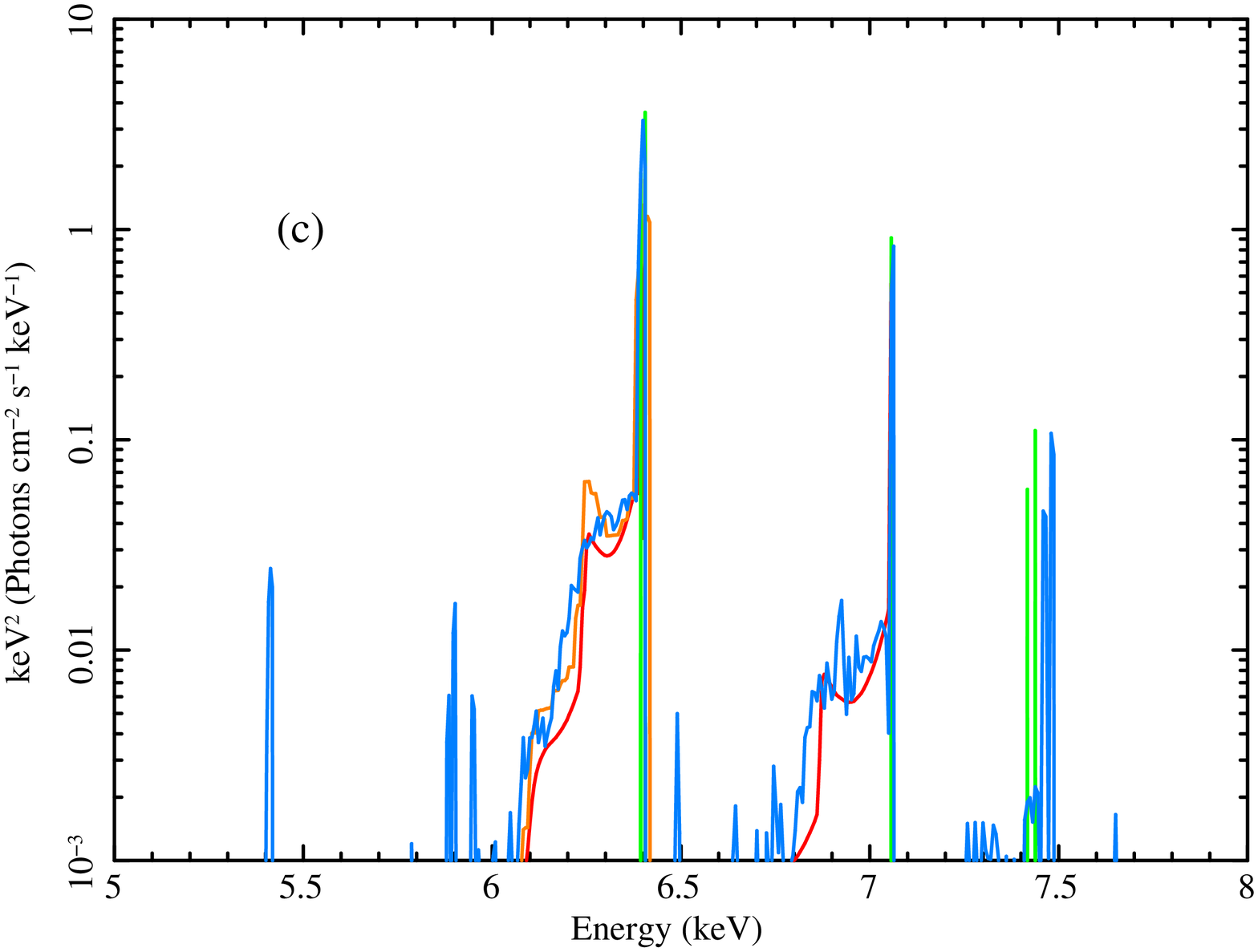}{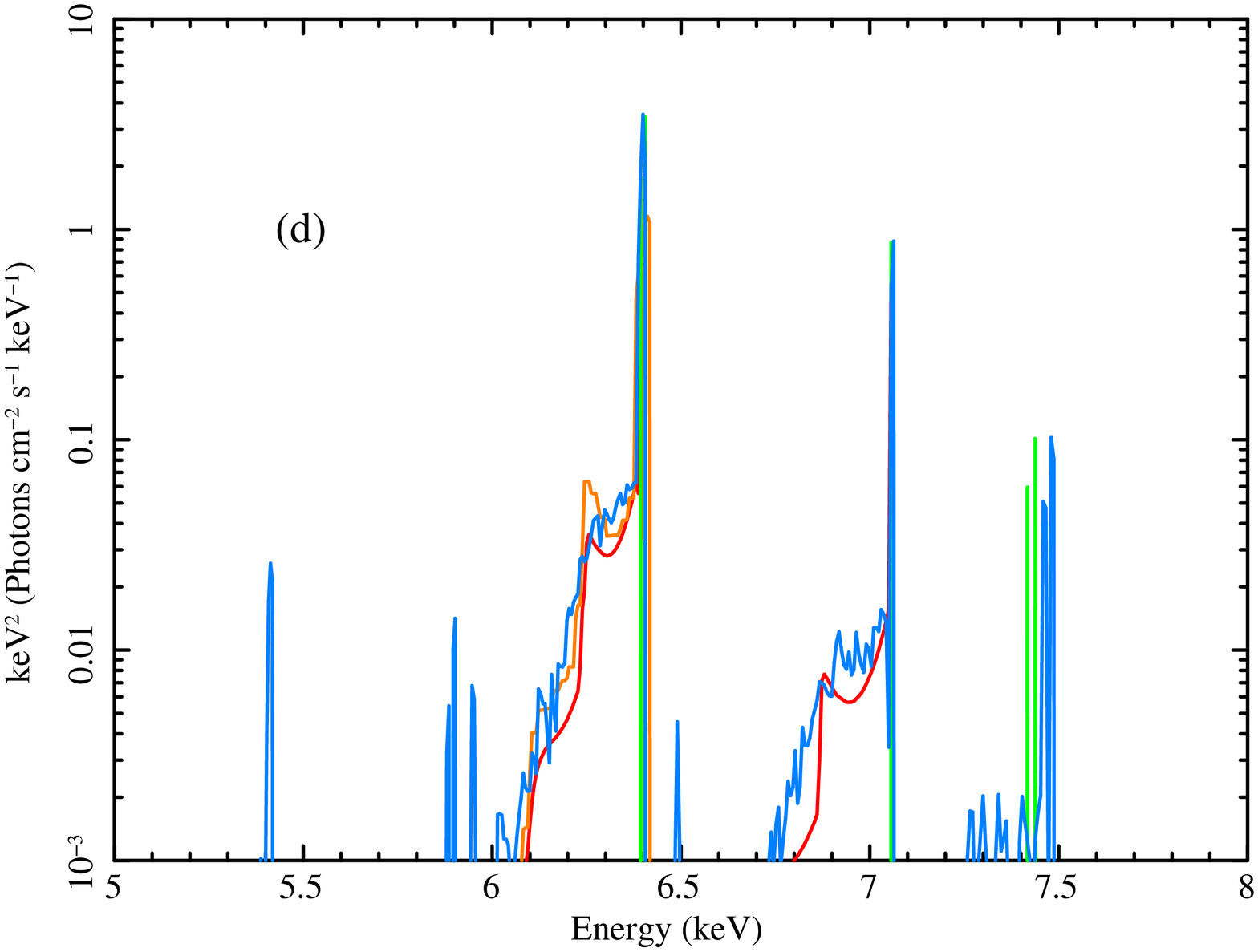}
\caption{Comparison of the emission lines among the torus models. Left: $N_{\mathrm{clump}}^{\mathrm{Equ}} = 5.0$. Right: $N_{\mathrm{clump}}^{\mathrm{Equ}} = 10.0$. Top: $i = 20.0$ degree. Bottom: $i = 87.0$ degree. Red line: MYTorus model. Orange line: Ikeda model. Green line: CTorus model. Blue line: XCLUMPY model. Torus parameters are as follows: MYTorus model: $N_{\mathrm{H}}^{\mathrm{Equ}} = 1.0 \times 10^{24}$ cm$^{-2}$, $\theta_{\mathrm{open}} = 60.0$ degree (fixed). Ikeda model: $N_{\mathrm{H}}^{\mathrm{Equ}} = 1.0 \times 10^{24}$ cm$^{-2}$, $\theta_{\mathrm{open}} = 60.0$ degree. CTorus model: $N_{\mathrm{H}}^{\mathrm{Equ}} = 1.5 \times 10^{24}$ cm$^{-2}$, $\theta_{\mathrm{open}} = 60.0$ degree (fixed). XCLUMPY model: $N_{\mathrm{H}}^{\mathrm{Equ}} = 1.0 \times 10^{24}$ cm$^{-2}$, $\sigma = 30.0$ degree.}
\end{figure*}

\section{Monte Carlo Simulation}
\subsection{Material Properties and Physical Processes}
The MONACO framework \citep{Odaka11, Odaka16} is utilized to perform our ray-tracing simulations where a clumpy torus is irradiated by X-rays from its central position. For simplicity, we assume that all matter in the torus is neutral cold gas. Any thermal motion of the gas is ignored. As the physical processes, we take into account photoelectric absorption, fluorescence line emission after it, and Compton scattering. We assume that all Compton scattering is made by electrons bound to atoms or molecules, not by free electrons, unlike in most of previous works. Although our assumption may not hold if there are ionized plasma in the torus, the difference would be only slight energy shifts of scattered X-rays by electron binding energies \citep{Odaka11, Odaka16}. This is not important except for high energy-resolution spectroscopy like that by a microcalorimeter \citep{Hitomi18}. We adopt the photoelectric cross sections compiled in the National Institute of Standards and Technology (NIST) database \citep{Schoonjans11}, and Solar abundances by \citet{Anders89}.

\subsection{Table Model}
We construct an X-ray spectral table model that can be directly applied to the observed data on the XSPEC package \citep{Arnaud96}. First, for each set of the torus parameters, we determine the locations of all clumps with randomization. We put clumps following Equation (1), and then randomly shift their three dimensional locations if the clumps overlap. For each geometry set, we perform Monte-Carlo simulations by generating 0.2 billion primary photons. The reflected spectra (continuum and fluorescence lines) are accumulated from all azimuth directions for each range of the inclination angle.

In each run, the photon energies are distributed according to a power law spectrum with a photon index of 2.0 in an energy range of 0.5--500 keV. The information on the initial energy of each input photon is recorded in the simulation. Hence, we are able to reproduce the spectra for various spectral parameters (photon index $\Gamma$ and high-energy cutoff $E_{\mathrm{cut}}$) by multiplying different weights to primary photon energies in the original simulation data. Table 1 also summarizes the range and grid values of the spectral parameters available in our table model.

\section{Results}
\subsection{Dependence of Reflected Continuum Spectrum on Torus Parameters}
We investigate the dependence of the reflected X-ray spectrum on the torus parameters. Here we adopt the following values as default (fixed) parameters unless otherwise stated: $\log N_{\mathrm{H}}/$cm$^{-2} = 24.0$, $\sigma = 40.0$ degree, $i = 60.0$ degree, $\Gamma = 2.0$, and $E_{\mathrm{cut}} = 100$ keV. Figure 2 shows the dependence of the X-ray spectrum on the (a) hydrogen column density along the equatorial plane, (b) torus angular width, and (c) inclination angle.

In Figure 2(a), we find that the continuum flux above $\sim$20 keV increases and the overall spectrum hardens with the hydrogen column density along the equatorial plane. This is because the intensity of the Compton-reflected continuum and the amount of self-absorption by the torus increase with the total mass of the torus, which is proportional to $N_{\mathrm{H}}$ when the other parameters are fixed. Figure 2(b) indicates that the X-ray flux below 7.1 keV (the K-edge of cold iron) and that above 7.1 keV decreases and increases with $\sigma$, respectively. The total mass and covering fraction of the reflector increase with $\sigma$ for a fixed $N_{\mathrm{H}}$, leading to stronger reflection and self-absorption by the torus. In Figure 2(c), we find that the flux below 20.0 keV decreases with the inclination angle. This is because the line-of-sight absorption of the reflection component increases with the viewing angle.

\subsection{Dependence of Fe K$\alpha$ line profile on Torus Parameters}
We investigate the dependence of Fe K$\alpha$ line profile on the torus parameters. We adopt the same default parameters as in Section 4.1 ($\log N_{\mathrm{H}}/$cm$^{-2} = 24.0$, $\sigma = 40.0$ degree, $i = 60.0$ degree, $\Gamma = 2.0$, and $E_{\mathrm{cut}} = 100$ keV). Figure 3 shows the dependence of the Fe K$\alpha$ on the (a) hydrogen column density along the equatorial plane, (b) torus angular width, and (c) inclination angle.

We also investigate the dependence of the equivalent width of the total Fe K$\alpha$ line, that of the Compton shoulder component, and the Compton shoulder fraction relative to the total line intensity. We define these quantities as follows:
\begin{align}
\mathrm{EW}_{\mathrm{K}\alpha}	& =\int_{6.086\mathrm{keV}}^{6.404\mathrm{keV}}\frac{L(E)}{R(E)}dE\\
\mathrm{EW}_{\mathrm{CS}}		& =\int_{6.086\mathrm{keV}}^{6.390\mathrm{keV}}\frac{L(E)}{R(E)}dE\\
f_{\mathrm{CS}}					& =\frac{\int_{6.086\mathrm{keV}}^{6.390\mathrm{keV}}L(E)dE}{\int_{6.086\mathrm{keV}}^{6.404\mathrm{keV}}L(E)dE}
\end{align}
where R(E) and L(E) are the spectra of the reflection continuum and emission line, respectively. We set the upper boundary of the integrals to 6.404 keV for $\mathrm{EW}_{\mathrm{K}\alpha}$, which is just above the Fe K$\alpha_1$ (6.403 keV), and to 6.390 keV for $\mathrm{EW}_{\mathrm{CS}}$, just below Fe K$\alpha_2$ (6.390 keV). The lower boundary of the integrals are set to be 6.086 keV, corresponding to the lowest energy of the second-order Compton shoulder of Fe K$\alpha_2$.

In Figure 4, we plot the dependence of the equivalent widths and the Compton shoulder fraction on the (a)-(b) hydrogen column density, (c)-(d) torus angular width, and (e)-(f) inclination angle.

As noticed from Figures 4 (a) and (c), the equivalent width of the Fe K$\alpha$ line increases with $N_{\mathrm{H}}$ and $\sigma$, confirming the same trends found with smooth torus models \citep{Ikeda09, Murphy09}. Figures 4 (b) and (d) show that the Compton shoulder fraction also increases with $N_{\mathrm{H}}$ and $\sigma$. This is because the probability that a fluorescent line is further Compton reflected by surrounding matter increases with the total mass of the torus. In Figure 4 (e) and (f), we find that the equivalent width and the Compton shoulder fraction little depend on the inclination angle.

\subsection{Comparison of Torus Models}
We compare our model with other torus models: MYTorus model \citep{Murphy09}, Ikeda model \citep{Ikeda09}, and CTorus model \citep{Liu14}. To consider similar geometry among the models as much as possible, we set the torus parameters as follows: for MYTorus model $N_{\mathrm{H}}^{\mathrm{Equ}} = 1.0 \times 10^{24}$ cm$^{-2}$ and $\theta_{\mathrm{open}} = 60.0$ degree (fixed in the model); for Ikeda model $N_{\mathrm{H}}^{\mathrm{Equ}} = 1.0 \times 10^{24}$ cm$^{-2}$, and $\theta_{\mathrm{open}} = 60.0$ degree; for CTorus model $N_{\mathrm{H}}^{\mathrm{Equ}} = 1.5 \times 10^{24}$ cm$^{-2}$ and $\theta_{\mathrm{open}} = 60.0$ degree (fixed); for XCLUMPY model $N_{\mathrm{H}}^{\mathrm{Equ}} = 1.0 \times 10^{24}$ cm$^{-2}$ and $\sigma = 30.0$ degree. We set $N_{\mathrm{H}}^{\mathrm{Equ}} = 1.5 \times 10^{24}$ cm$^{-2}$ in CTorus model because \cite{Liu14} define it as the value when all clumps are exactly aligned along the radial direction, whereas our definition refers to the case where the clumps are randomly distributed (see Equation (5)).

Figure 5 shows the comparison of the reflected X-ray continuum for (a) $N_{\mathrm{clump}}^{\mathrm{Equ}} = 5.0$ (this parameter is relevant only for CTorus and XCLUMPY) and $i = 20.0$ degree, (b) $N_{\mathrm{clump}}^{\mathrm{Equ}} = 10.0$ and $i = 20.0$ degree, (c) $N_{\mathrm{clump}}^{\mathrm{Equ}} = 5.0$ and $i = 87.0$ degree, and (d) $N_{\mathrm{clump}}^{\mathrm{Equ}} = 10.0$ and $i = 87.0$ degree. In Figures 5 (c) and (d) (i.e., edge-on), we find that the fluxes above 20 keV are almost the same among all the models, whereas those below 20 keV in the clumpy torus models (CTorus and XCLUMPY) are larger than those in the smooth torus models (MYTorus and Ikeda torus). This is mainly because a significant fraction of photons reflected by the far-side torus can reach the observer without being absorbed by the near-side torus in clumpy geometry in the case of edge-on view. This effect is more prominent in XCLUMPY than in CTorus. In CTorus, the clumps are confined within an elevation angle of 30 degree from the equatorial plane with a constant number density, whereas in XCLUMPY they are distributed to higher elevation angles with decreasing number densities. In XCLUMPY, photons can be reflected at higher elevation angles, which are subject to smaller line-of-sight absorption in edge-on view, thus producing larger soft X-ray fluxes, than in CTorus. As we mentioned in Section 1, smooth torus models cannot explain a large amount of the unabsorbed reflection component often seen in the X-ray spectra of heavily obscured AGNs \citep{Tanimoto16, Tanimoto18}. This feature can be naturally explained with the clumpy torus models.

In Figures 5 (a) and (b) (i.e., face-on), we find that Ikeda model, which adopts a similar spherical torus geometry to those in CTorus and XCLUMPY, produces higher fluxes at energies below several keV than the clumpy torus models. The trend is opposite to the edge-on case; when viewed edge-on, in Ikeda model the soft X-ray reflection component comes from the entire surface of the smooth torus without being absorbed, whereas in the clumpy torus models it comes from clumps at various depths (as measured from the observer) and hence is subject to absorption by other clumps while traveling inside the torus. We note that CTorus and XCLUMPY produce almost the same fluxes below 20 keV, unlike in the edge-on case. This is because the averaged column density along the line-of-sight responsible for the self absorption  is similar between the two models, even if the clump distribution in XCLUMPY is spread wider in elevation angles than in CTorus. In the torus geometry of MYTorus (bagel-like shape), the area of the irradiated surface is smaller and hence the soft X-ray flux of the reflection component becomes weaker than in Ikeda model.

Figure 6 shows the comparison of the emission lines for (a) $N_{\mathrm{clump}}^{\mathrm{Equ}} = 5.0$ and $i = 20.0$ degree, (b) $N_{\mathrm{clump}}^{\mathrm{Equ}} = 10.0$ and $i = 20.0$ degree, (c) $N_{\mathrm{clump}}^{\mathrm{Equ}} = 5.0$ and $i = 87.0$ degree, and (d) $N_{\mathrm{clump}}^{\mathrm{Equ}} = 10.0$ and $i = 87.0$ degree. We note that there are limitations in the treatment of fluorescence lines in the previous models. MYTorus model includes only Fe K$\alpha$ and Fe K$\beta$ (with the Compton shoulder), and Ikeda model only Fe K$\alpha$ (with the Compton shoulder). CTorus model includes other lines than Fe K$\alpha$ and K$\beta$ (such as Ni K$\alpha$, K$\beta$), but the Compton shoulders are not taken into account. The XCLUMPY model includes all prominent fluorescence lines from many elements with the Compton shoulder It should be also stressed that XCLUMPY accurately calculates the smeared profiles of the Compton shoulders since the MONACO framework considers atomic and molecular binding of electrons responsible for the scattering \citep{Odaka16}.

\begin{deluxetable*}{lllllll}
\tablecaption{Torus Structure from infrared observation}
\tablehead
{
$R_{\mathrm{inner}}$			& $R_{\mathrm{outer}}$		    & $N_{\Clu}^{\Equ}$				& $\tau_{\mathrm{v}}$			&
$\sigma$						& $i$						    & $\log L_{\Bol}/$erg s$^{-1}$	                                \\
(1)							    & (2)							& (3)						    & (4)							&
(5)							    & (6)							& (7)}
\startdata
$0.071_{-0.003}^{+0.003}$		& $1.4_{-0.2}^{+0.3}$		    & $7_{-1}^{+1}$					& $37_{-2}^{+3}$				&
$65_{-5}^{+2}$			    	& $63_{-2}^{+4}$			    & $43.5_{-0.1}^{+1.1}$
\enddata
\tablecomments{Column (1): Inner radius of the torus in units of pc. Column (2): Outer radius of the torus in units of pc. Column (3): number of the clump along the equatorial plane. Column (4): optical depth of each cloud. Column (5): torus angular width in units of degree. Column (6): inclination angle in units of degree. Column (7): logarithmic bolometric luminosity.}
\end{deluxetable*}
\begin{deluxetable*}{lllllll}
\tablecaption{Summary of Observations}
\tablehead
{
Observatory			    		& Observation ID			    & Start Date					& End Date                      &
Exposure						& Nominal Position			    & Reference						                                \\
(1)							    & (2)							& (3)						    & (4)							&
(5)							    & (6)							& (7)}
\startdata
\textit{Suzaku}		    		& 701036010					    & 2006 Jul 21 					& 2006 Jul 23					&
108	    						& XIS						    & (1)							                                \\
\textit{XMM-Newton}		    	& 0701981001				    & 2013 Feb 03					& 2013 Feb 03					&
59  							& \nodata					    & (2)							                                \\
\textit{NuSTAR}				    & 30002038004				    & 2013 Feb 03					& 2013 Feb 03				    &
40		    					& \nodata					    & (2)                          
\enddata
\tablecomments{Column (1): observatory. Column (2): observation identification number. Column (3): start date in units of ymd. Column (4): end date in units of ymd. Column (5): exposure time in units of ksec. Column (6): nominal position. (7) reference of the previous work.}
\tablerefs{(1) \cite{Yang09}. (2) \cite{Arevalo14}.}
\end{deluxetable*}

\section{Application to Circinus galaxy}
We apply the XCLUMPY model to broadband X-ray spectra of the Circinus galaxy. The Circinus galaxy ($z=0.0014$) is one of the closest (4.2 Mpc: \citealt{Freeman77}) obscured AGNs and is an ideal target for investigating torus structure. \cite{Ichikawa15} applied the CLUMPY model to its infrared data and derived the torus parameters (Table 2). Table 3 summarizes the X-ray observations of the Circinus galaxy analyzed in this paper. \textit{Suzaku} observed this object in 2006 July. Simultaneous observations with \textit{XMM-Newton} and \textit{NuSTAR} were performed in 2013 February.

\subsection{Data Analysis}
\subsubsection{\textit{Suzaku}}
\textit{Suzaku} \citep{Mitsuda07} is the fifth Japanese X-ray astronomical satellite, which was in operation between 2005 and 2015. It carries four X-ray CCD cameras called the X-ray imaging spectrometers (XIS: \citealt{Koyama07}) and a non imaging, collimated hard X-ray instrument called the hard X-ray detector (HXD: \citealt{Takahashi07}). XIS0, XIS2, and XIS3 are front-side-illuminated CCDs (FIXIS: 0.4--12.0 keV) and XIS1 is the back-illuminated one (BIXIS: 0.2--12.0 keV). The HXD consists of two types of detectors: the PIN photo-diodes (PIN: 10--70 keV) and gadolinium silicon oxide (GSO: 40--600 keV) scintillation counters \citep{Kokubun07}. We analyzed the XIS and HXD data of the Circinus galaxy, using the HEAsoft 6.24 and the calibration database (CALDB) released on 2016 June 7 (XIS) and 2011 September 13 (HXD). We reprocessed the unfiltered XIS and HXD data by using \textsf{aepipeline}.

To extract the XIS light curves and spectra, we accumulated photon events in the circle of a 1 arcmin radius centered on the source peak, by subtracting the background taken from a source-free circular region of a 1 arcmin radius. We generated the XIS redistribution matrix files (RMF) by using \textsf{xisrmfgen} and ancillary response files (ARF) by using \textsf{xissimarfgen} \citep{Ishisaki07}. We combined the source spectra, background spectra, RMF, and ARF of FIXIS by using \textsf{addascaspec}. We binned the BIXIS and FIXIS spectra to contain at least 50 counts per bin. We generated the light curves and spectra of HXD-PIN by using \textsf{hxdpinxblc} and \textsf{hxdpinxbpi} and HXD-GSO by using \textsf{hxdgsoxblc} and \textsf{hxdgsoxbpi}. We utilized the tuned background files \citep{Fukazawa09} to reproduce the spectra of non X-ray background (NXB). The spectrum of the cosmic X-ray background (CXB) simulated with the energy response for diffuse emission was added to the NXB spectrum of HXD-PIN, whereas the CXB is ignored for HXD-GSO.

\subsubsection{\textit{XMM-Newton}}
\textit{XMM-Newton} \citep{Jansen01} is the second ESA X-ray astronomical satellite. It carries three X-ray CCD cameras: one EPN \citep{Struder01} and two MOS \citep{Turner01}. We analyzed the EPN and MOS data using the science analysis software (SAS) 16.10 and current calibration file (CCF) released on 2017 December 15. We reprocessed the unfiltered EPN and MOS data by using \textsf{epproc} and \textsf{emproc}, respectively. We extracted the source spectra from the circle of a 1 arcmin radius centered on the flux peak. The background was taken from a source-free circular region of 1 arcmin radius in the same CCD chip. We generated the RMF by using \textsf{rmfgen} and ARF by using \textsf{arfgen}. We combined the source spectra, background spectra, RMF, and ARF of MOS by using
\textsf{addascaspec}. We then binned the EPN and MOS spectra to contain at least 50 counts per bin.

\subsubsection{\textit{NuSTAR}}
\textit{NuSTAR} \citep{Harrison13} is the first astronomical satellite capable of focusing hard X-rays above 10 keV. It carries two co-aligned grazing incidence telescopes coupled with two focal plane modules (FPMs: FPMA and FPMB), which cover an energy range of 3--79 keV. We analyzed the FPM data, using the HEAsoft 6.24 and CALDB released on 2018 April 19. Utilizing \textsf{nupipeline} and \textsf{nuproducts}, we extracted the spectrum from the 1-arcmin radius circle centered at the source peak and subtracted the background from a source-free circular region of a 1 arcmin radius. We combined the source spectra, background spectra, RMF, and ARF by using \textsf{addascaspec}. The combined spectrum is then binned to contain at least 50 counts per bin.

\begin{figure*}
\plottwo{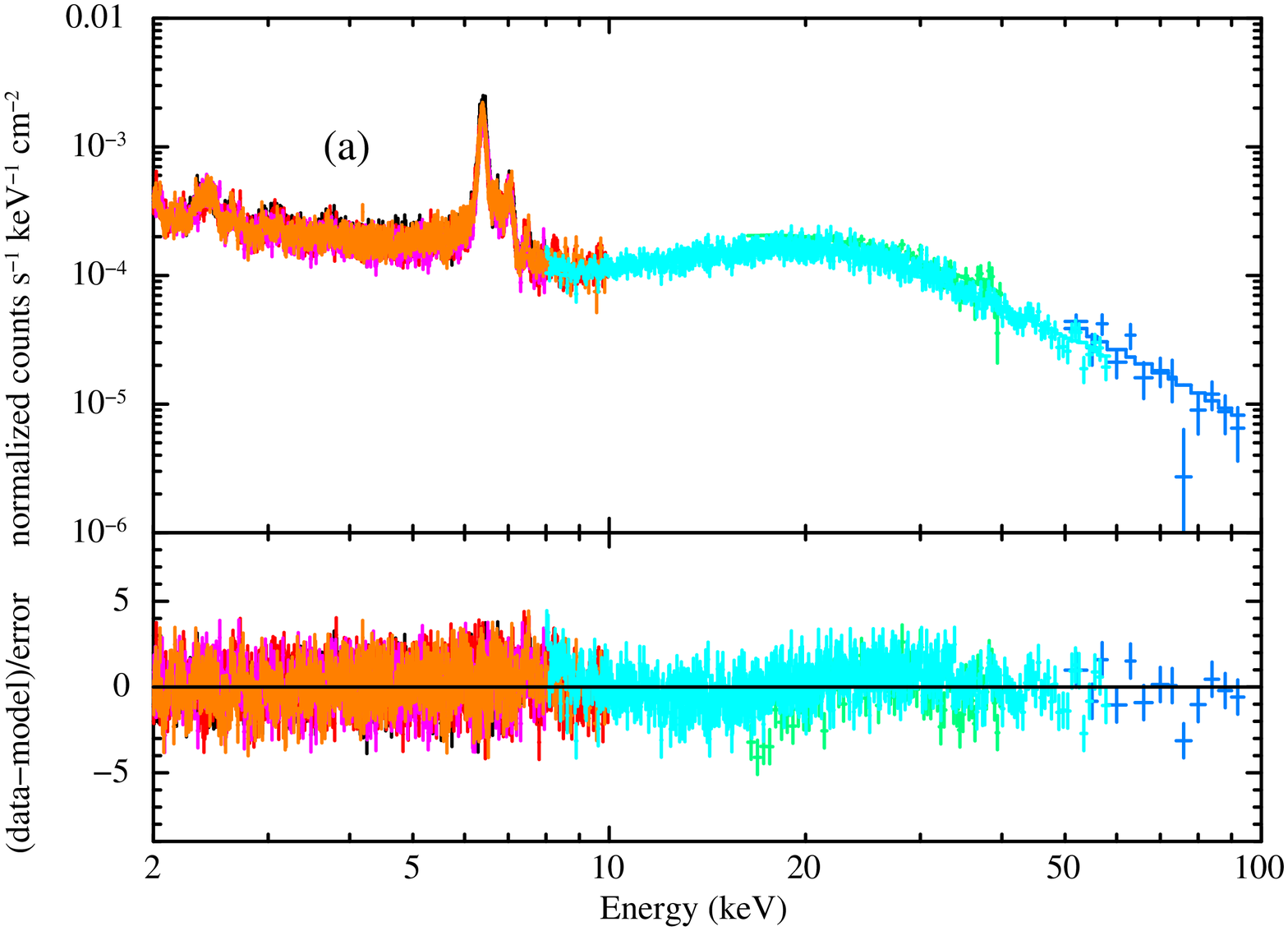}{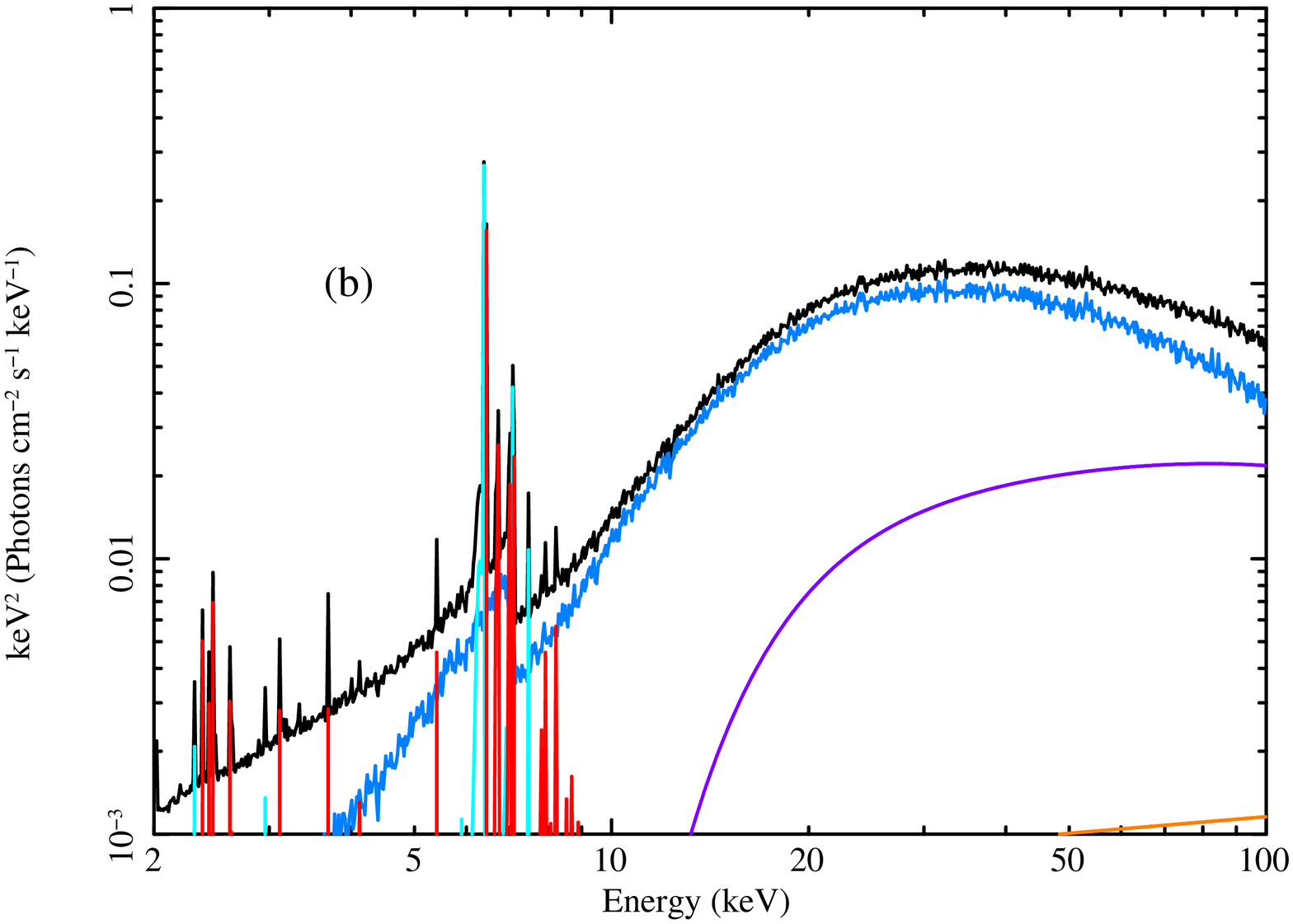}
\caption{Left: folded X-ray spectra fitted with the XCLUMPY model. Right: best-fitting model. Left: Black crosses: \textit{Suzaku}/BIXIS. Red crosses: \textit{Suzaku}/FIXIS. Green crosses: \textit{Suzaku}/PIN. Blue crosses: \textit{Suzaku}/GSO. Pink crosses: \textit{XMM-Newton}/EPN. Orange crosses: \textit{XMM-Newton}/MOS. Light Blue crosses: \textit{NuSTAR}/FPMs. Upper panel: solid curves represent the best-fitting model. Lower panel: each crosses represents residual. Right: Black line: total. Purple line: direct component. Blue line: reflection component from the torus. Light Blue lines: emission line from the torus. Orange line: contamination from the CGX1. Red lines: contamination from the CGX2.}
\end{figure*}
\begin{deluxetable*}{llllllllll}
\tablecaption{Best-fit Parameters with the XCLUMPY model}
\tablehead
{                      
$C_{\mathrm{BIXIS}}$			& $C_{\mathrm{EPN}}$			& $C_{\mathrm{MOS}}$	        & $C_{\mathrm{FPM}}$	        & $N_{\Hyd}^{\Los}$				&
$N_{\Hyd}^{\Equ}$				& $\sigma$                      & $i$                           & $\Gamma$                      & $\chi^2/$dof                 \\
(01)						    & (02)							& (03)							& (04)					    	& (05)						    &
(06)							& (07)							& (08)                          & (09)                          &                              \\                      	
$E_{\mathrm{cut}}$	    		& $N_{\mathrm{Dir}}$	        & $f_{\mathrm{scat}}$	    	& $E_{\mathrm{FeK}\alpha}$      & $N_{\mathrm{FeK}\alpha}$      &
$E_{\mathrm{FeK}\beta}$         & $N_{\mathrm{FeK}\beta}$       & $\log L_{2-10}$               & $\log L_{10-50}$              & reduced $\chi^2$              \\
(10)							& (11)							& (12)	        				& (13)                          & (14)						    &
(15)							& (16)							& (17)                          & (18)}
\startdata
$1.08_{-0.01}^{+0.01}$  		& $0.96_{-0.01}^{+0.01}$		& $1.02_{-0.01}^{+0.01}$	    & $0.88_{-0.01}^{+0.01}$	    & $4.86_{-0.04}^{+0.07}$	    &
$9.08_{-0.08}^{+0.14}$	    	& $14.7_{-0.39}^{+0.44}$        & $78.3_{-0.15}^{+0.17}$        & $1.80_{-0.03}^{+0.01}$        & $3646/2989$                   \\
$ 114_{-8.91}^{+9.57}$          & $0.41_{-0.01}^{+0.01}$        & $0.18_{-0.01}^{+0.01}$        & $6.43_{-0.01}^{+0.01}$	    & $1.07_{-0.01}^{+0.01}$		&
$7.12_{-0.01}^{+0.01}$          & $0.13_{-0.01}^{+0.01}$        & $42.8$                        & $42.8$                        & $1.22$
\enddata
\tablecomments{Column (01): cross-calibration constant of the \textit{Suzaku}/BIXIS relative to the \textit{Suzaku}/FIXIS. Column (02): cross-calibration constant of the \textit{XMM-Newton}/EPN relative to the \textit{Suzaku}/FIXIS. Column (03): cross-calibration constant of the \textit{XMM-Newton}/MOS relative to the \textit{Suzaku}/FIXIS. Column (04): cross-calibration constant of the \textit{NuSTAR}/FPM relative to the \textit{Suzaku}/FIXIS. Column (05): hydrogen column density along the line of sight in units of 10$^{24}$ cm$^{-2}$. Column (06): hydrogen column density along the equatorial plane in units of 10$^{24}$ cm$^{-2}$. Column (07): torus angular width in units of degree. Column (08): inclination angle in units of degree. Column (09): photon index. Column (10): cutoff energy in units of keV. Column (11): normalization of the direct component in units of photons keV$^{-1}$ cm$^{-2}$ s$^{-1}$. Column (12): scattering fraction in units of percent. Column (13): energy of the additional Fe K$\alpha$ emission line. Column (14): normalization of the additional Fe K$\alpha$ emission line in units of 10$^{-4}$ photons cm$^{2}$ s$^{-1}$. Column (15): energy of the additional Fe K$\beta$ emission line. Column (16): normalization of the additional Fe K$\beta$ emission line in units of 10$^{-4}$ photons cm$^{2}$ s$^{-1}$. Column (17): logarithmic intrinsic luminosity in the 2--10 keV. Column (18): logarithmic intrinsic luminosity in the 10--50 keV.}
\end{deluxetable*}

\subsection{Spectral Analysis}
Since our main interest is the reflection component from the torus, here we only analyze the spectra above 2 keV, in order to avoid complexity in modelling the soft X-ray emission \citep{Matt96}. To best constrain the torus parameters, we perform simultaneous fit to the \textit{Suzaku}/BIXIS (2--8 keV), \textit{Suzaku}/FIXIS (2--10 keV), \textit{Suzaku}/PIN (16-40 keV), \textit{Suzaku}/GSO (50--100 keV), \textit{XMM-Newton}/EPN (2--8 keV), \textit{XMM-Newton}/MOS (2--10 keV), and \textit{NuSTAR}/FPM (8--60 keV). Possible time variability between the two epochs (2006 and 2013) is ignored, which is found to be not required from the data. We apply the XCLUMPY model to reproduce the torus reflection component. The whole model is represented as follows in the XSPEC \citep{Arnaud96} terminology:
\begin{align}
& \mathsf{const1*phabs}\nonumber\\
& *(\mathsf{zphabs*cabs*zcutoffpl+const2*zcutoffpl}\nonumber\\
& + \mathsf{atable\{xclumpy\_R.fits\}})\nonumber\\
& + \mathsf{atable\{xclumpy\_L.fits\}+zgauss+zgauss}\nonumber\\
& + \mathsf{phabs*powerlw+phabs*(apec+phabs*mekals))}\nonumber\\
\end{align}

Below we explain the details of each component:
\begin{enumerate}
\item Cross-normalization factor (const1). We multiply a constant to take into account the difference in the absolute flux calibration among the instruments. We set this value of \textit{Suzaku}/FIXIS unity as a reference. The cross-normalizations of \textit{Suzaku}/BIXIS ($N_{\mathrm{BIXIS}}$), \textit{XMM-Newton}/EPN ($N_{\mathrm{EPN}}$), \textit{XMM-Newton}/MOS ($N_{\mathrm{MOS}}$), and \textit{NuSTAR}/FPM ($N_{\mathrm{FPM}}$) are left as a free parameter. The cross-normalization of \textit{Suzaku}/PIN and \textit{Suzaku}/GSO is set to 1.16, according to the calibration results obtained using the Crab Nebula.
\item Galactic absorption (phabs). The hydrogen column density is fixed at $0.525 \times 10^{22}$ cm$^{-2}$, a value estimated from the H$_{\mathrm{I}}$ map \citep{Kalberla15}.
\item Transmitted component through the torus, which is subject to photoelectric absorption (zphabs) and Compton scattering (cabs). The line-of-sight column density is determined by the torus parameters (see Equation 3). We model the intrinsic continuum by a power-law with an exponential cutoff (zcutoffpl).
\item Unabsorbed scattered component. The scattering fraction $f_{\mathrm{scat}}$ (const2) is multiplied to the same model as the intrinsic continuum by linking the photon index, cutoff energy, and normalization.
\item Reflection continuum (xclumpy\_R.fits) from the torus, based on the XCLUMPY model. The photon index, cutoff energy, and normalization are linked to those of the intrinsic continuum.
\item Fluorescence lines (xclumpy\_L.fits) from the torus, based on the XCLUMPY model. The photon index, cutoff energy,  and normalization are linked to those of the intrinsic continuum. Since we find that the XCLUMPY model slightly underestimates the observed line fluxes, we add additional two gaussians for Fe K$\alpha$ and Fe K$\beta$. The cause of the discrepancy is not clear: there might be diffuse Fe-K emission not originating from the torus \citep{Marinucci13}, or the metal abundance is higher than the Solar value \citep{Hikitani18}.
\item Contamination from CGX1 (X-ray binary). We adopt the same model and flux as in \cite{Arevalo14}, who examined the spectrum of CGX1 with \textit{Chandra}.
\item Contamination from CGX2 (supernova remnant). We also assume the same model and flux as in \cite{Arevalo14}.
\end{enumerate}

\subsection{Results and Discussion}
Our model can well reproduce the broadband X-ray spectra observed with the three satellites. Table 4 summarizes the best-fit parameters. Figure 7 shows the (a) folded X-ray spectra and (b) best-fit models. We confirm that the torus of the Circinus galaxy is heavily Compton-thick; the column density in the line-of-sight and that along the equatorial plane are estimated to be $4.86_{-0.04}^{+0.07} \times 10^{24}$ cm$^{-2}$ and $9.08_{-0.08}^{+0.14} \times 10^{24}$ cm$^{-2}$, respectively. Accordingly, the X-ray spectrum is dominated by the reflection component rather than the transmitted one.

\subsubsection{Comparison to Previous Researches}
Here we compare our results with the previous X-ray works using the same data \citep{Yang09, Arevalo14}. \cite{Yang09} analyzed the \textit{Suzaku} spectra with the pexrav model, and obtained a line-of-sight absorption of $N_{\mathrm{H}}^{\mathrm{LOS}} = 4.70_{-0.32}^{+0.50} \times 10^{24}$ cm$^{-2}$ and a photon index of $\Gamma = 1.58_{-0.10}^{+0.07}$. The simultaneous \textit{XMM-Newton} and \textit{NuSTAR} data in 2013 were analyzed by \cite{Arevalo14}. Applying the MYTorus model, they obtained $N_{\mathrm{H}}^{\mathrm{LOS}} = 6.6_{-0.9}^{+0.9} \times 10^{24}$ cm$^{-2}$ and $\Gamma = 2.19_{-0.02}^{+0.02}$. Our result of the line-of-sight column density ($N_{\mathrm{H}}^{\mathrm{LOS}} = 4.86_{-0.04}^{+0.07} \times 10^{24}$ cm$^{-2}$) is consistent with both of them. The photon index ($\Gamma = 1.77_{-0.03}^{+0.06}$) is smaller than the \cite{Arevalo14} result, however. We infer that this is because the XCLUMPY model contains a larger flux of the unabsorbed reflection component, leading to a flatter intrinsic slope, compared with the MYTorus model case.

\subsubsection{Comparison of Torus Parameters by X-ray and infrared spectra}
It is interesting to compare the torus parameters obtained from the X-ray data (Table 4) with those from the infrared data (Table 2). The total V-band optical depth in the equatorial plane obtained with the CLUMPY model is $\tau_{\mathrm{V}} = 173_{-7}^{+26}$, which can be converted to $N_{\mathrm{H}}^{\mathrm{Equ}} = 0.35_{-0.05}^{+0.01} \times 10^{24}$ cm$^{-2}$ by assuming the gas-to-dust ratio in the Galaxy (\citealt{Draine03}: $N_{\mathrm{H}}/A_{\mathrm{V}} = 1.87 \times 10^{21}$ cm$^{-2}$ mag$^{-1}$). This value is much (by a factor of $\sim$ 26) smaller than the X-ray result, $N_{\mathrm{H}}^{\mathrm{Equ}} = 9.08_{-0.08}^{+0.14} \times 10^{24}$ cm$^{-2}$. The same conclusion holds when we compare the line-of-sight column density. This trend is consistent with the results of \cite{Burtscher15, Burtscher16}, who estimated $A_{\mathrm{V}}$ from the near-to-mid infrared colors. The torus angular width obtained from the infrared data ($\sigma = 65_{-5}^{+2}$ degree) is much larger than the X-ray results ($\sigma = 14.7_{-0.39}^{+0.44}$ degree). These results may be explained by the presence of a dusty polar outflow observed in the infrared interfere-metric observations \citep{Tristram14, Stalevski17}, which apparently makes the distribution of dust wider than that of gas. Since the extended outflow is subject to smaller extinction than that toward the very center of the nucleus, it works to reduce the observed (averaged) $A_{\mathrm{V}}$ value.

\section{Conclusion}
\begin{enumerate}
\item We have constructed the XCLUMPY model, a spectral model of X-ray reflection from the clumpy torus in an AGN, where the same torus geometry as in the infrared CLUMPY model is adopted. This enables us to directly compare the X-ray and infrared results, which trace the distribution of all matter and dust, respectively.
\item We found that the equivalent width of Fe K$\alpha$ line and the Compton shoulder fraction both  increase with the hydrogen column density along the equatorial plane and torus angular width.
\item Compared with smooth torus models, our model predicts a higher fraction of unabsorbed         reflection components as observed in many obscured AGNs.
\item Our model well reproduces the broadband X-ray spectra of the Circinus galaxy observed with \textit{XMM-Newton}, \textit{Suzaku}, and \textit{NuSTAR}. We confirm that the torus is heavily Compton thick and the spectrum is dominated by the reflection component from the torus.
\item In the Circinus galaxy, the column density obtained from the X-ray data is $>20$ times larger than that from the infrared data by assuming a Galactic gas-to-dust ratio. The torus angular width derived from X-rays is much smaller than that in the infrared band. This may be explained by the presence of a dusty polar outflow.
\end{enumerate}

\acknowledgements
Part of this work was financially supported by the Grant-in-Aid for JSPS fellows for young researchers (A.T.), and for Scientific Research 17K05384 (Y.U.) and 18H05861 (H.O.). This research has made use of data and/or software provided by the High Energy Astrophysics Science Archive Research Center (HEASARC), which is a service of the Astrophysics Science Division at NASA/GSFC and the High Energy Astrophysics Division of the Smithsonian Astrophysical Observatory. This research has also made use of the NASA/IPAC Extragalactic Database (NED), which is operated by the Jet Propulsion Laboratory, California Institute of Technology, under contract with the National Aeronautics and Space Administration. \facilities{\textit{XMM-Newton} (0701981001), \textit{Suzaku} (701036010), \textit{NuSTAR} (30002038004)} \software{HEAsoft 6.24 (HEASARC 2014), MONACO \citep{Odaka11, Odaka16}, SAS 17.00 \citep{Gabriel04}, XSPEC \citep{Arnaud96}}

\begin{figure*}
\plottwo{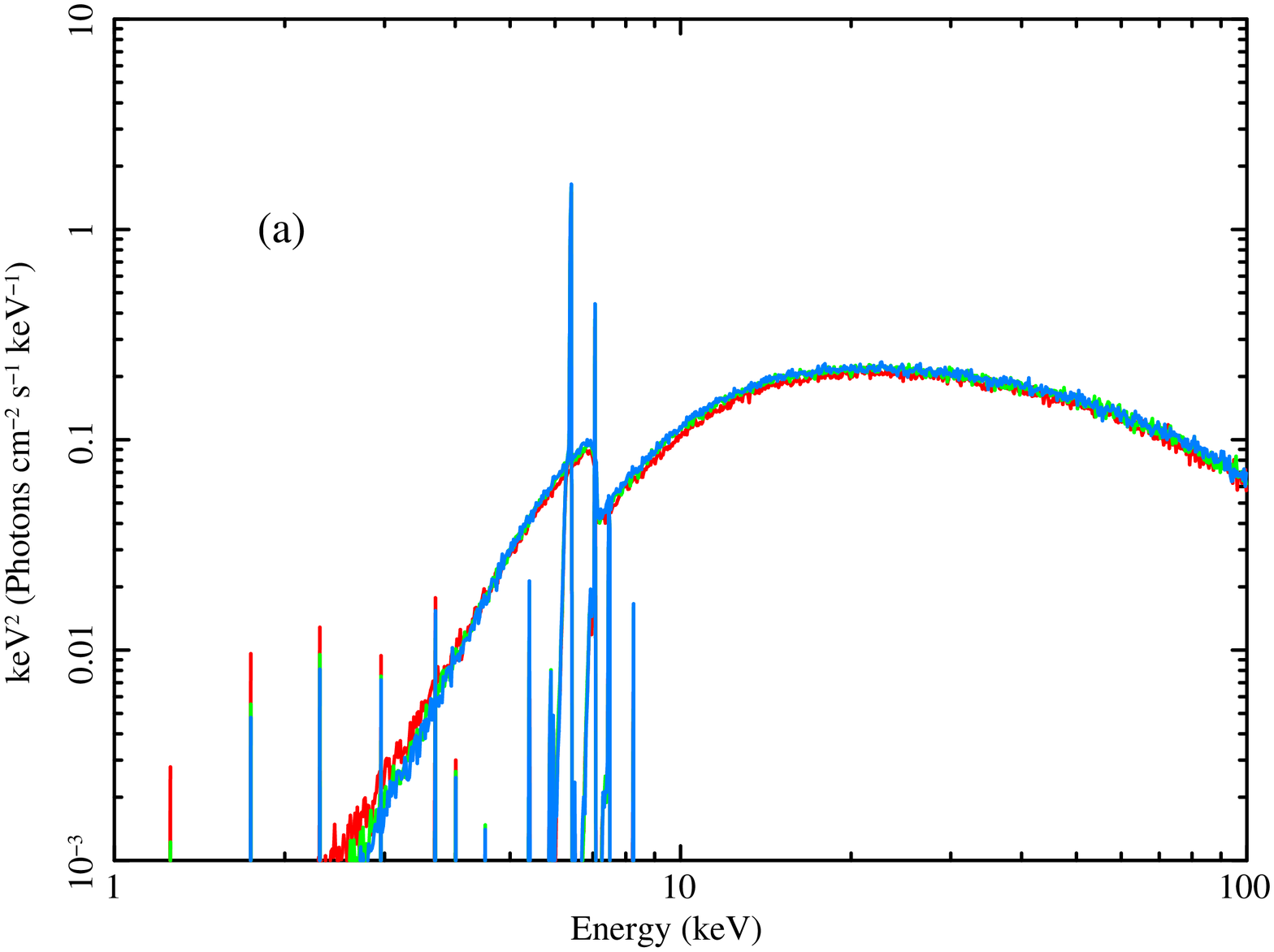}{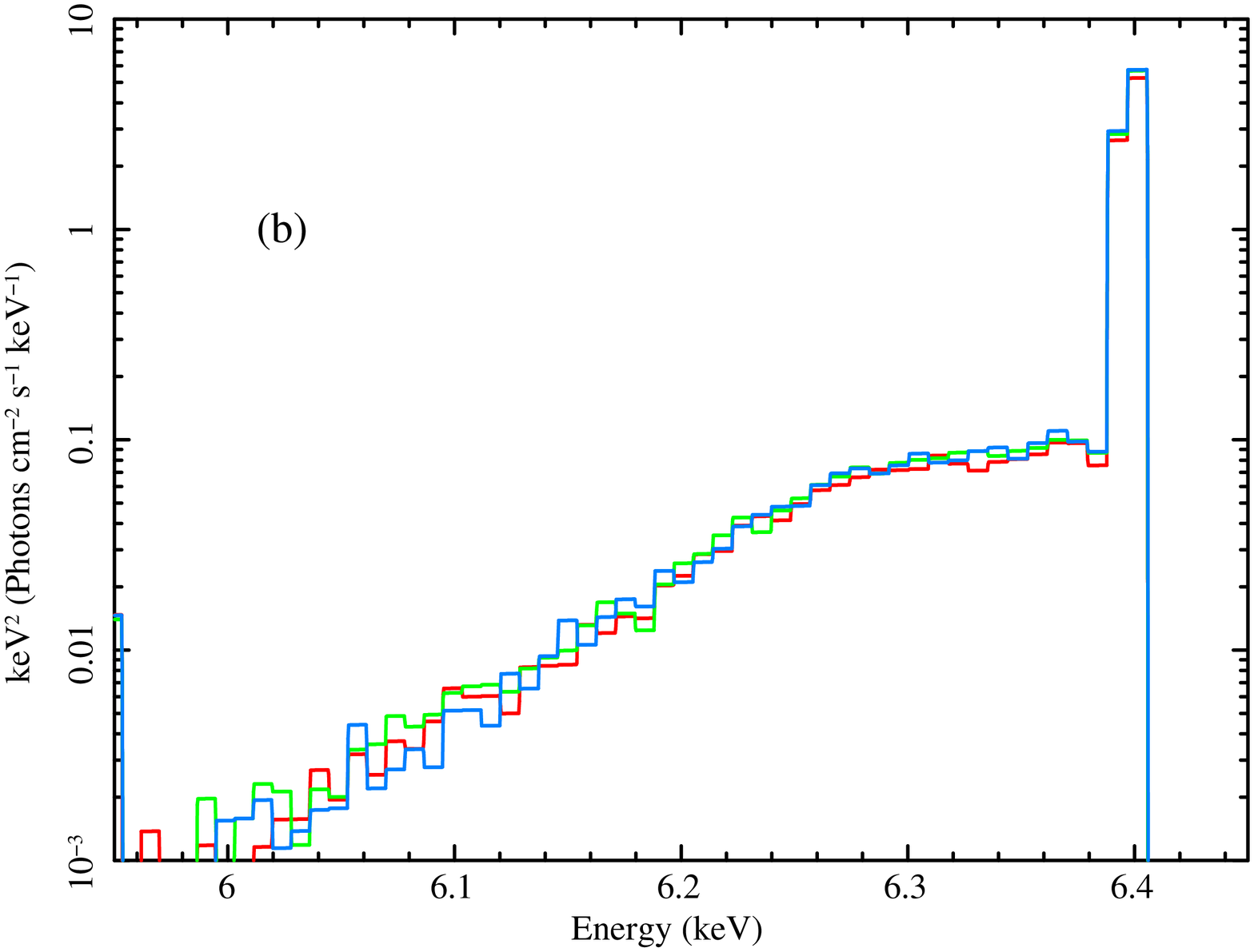}
\plottwo{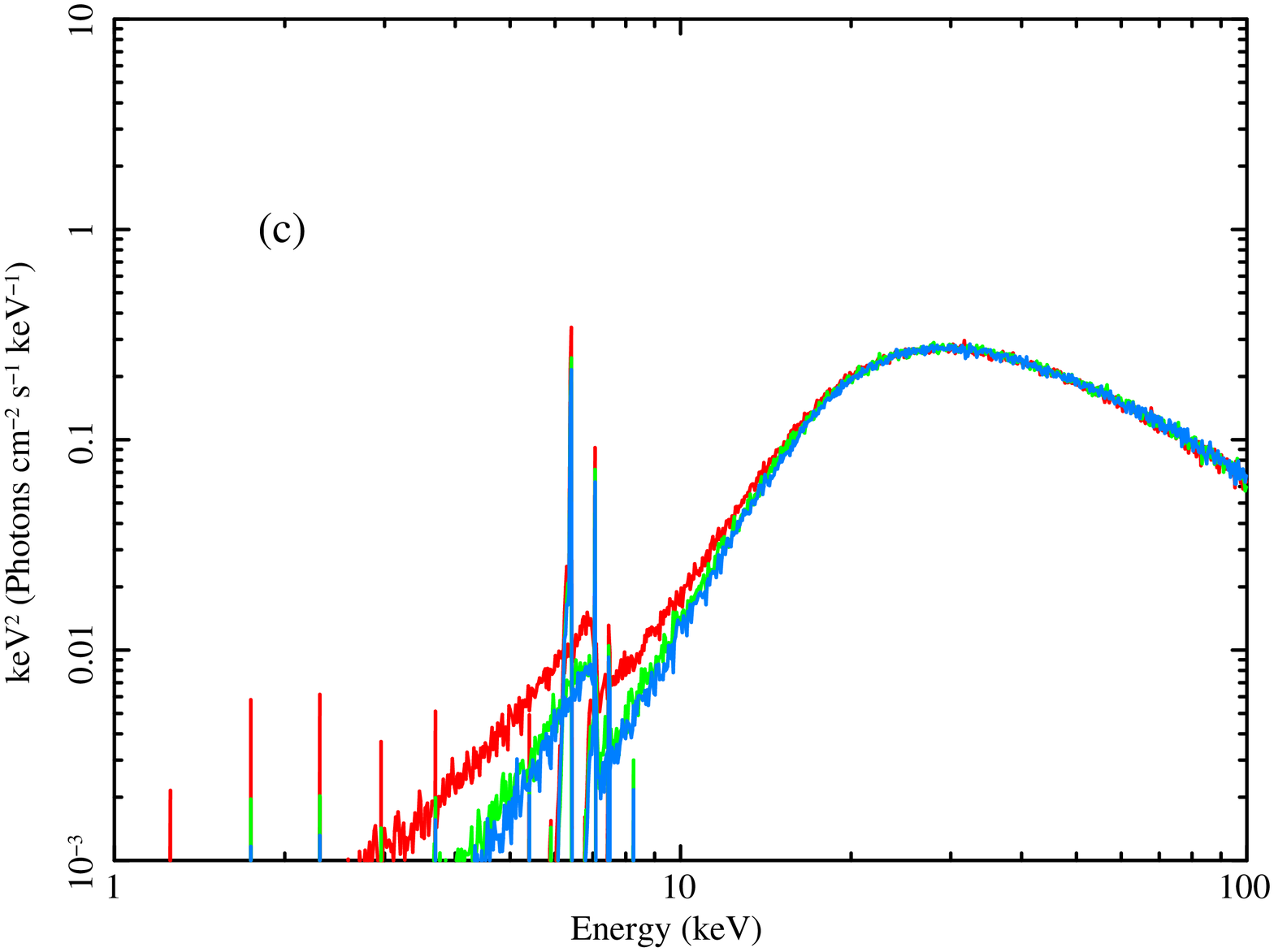}{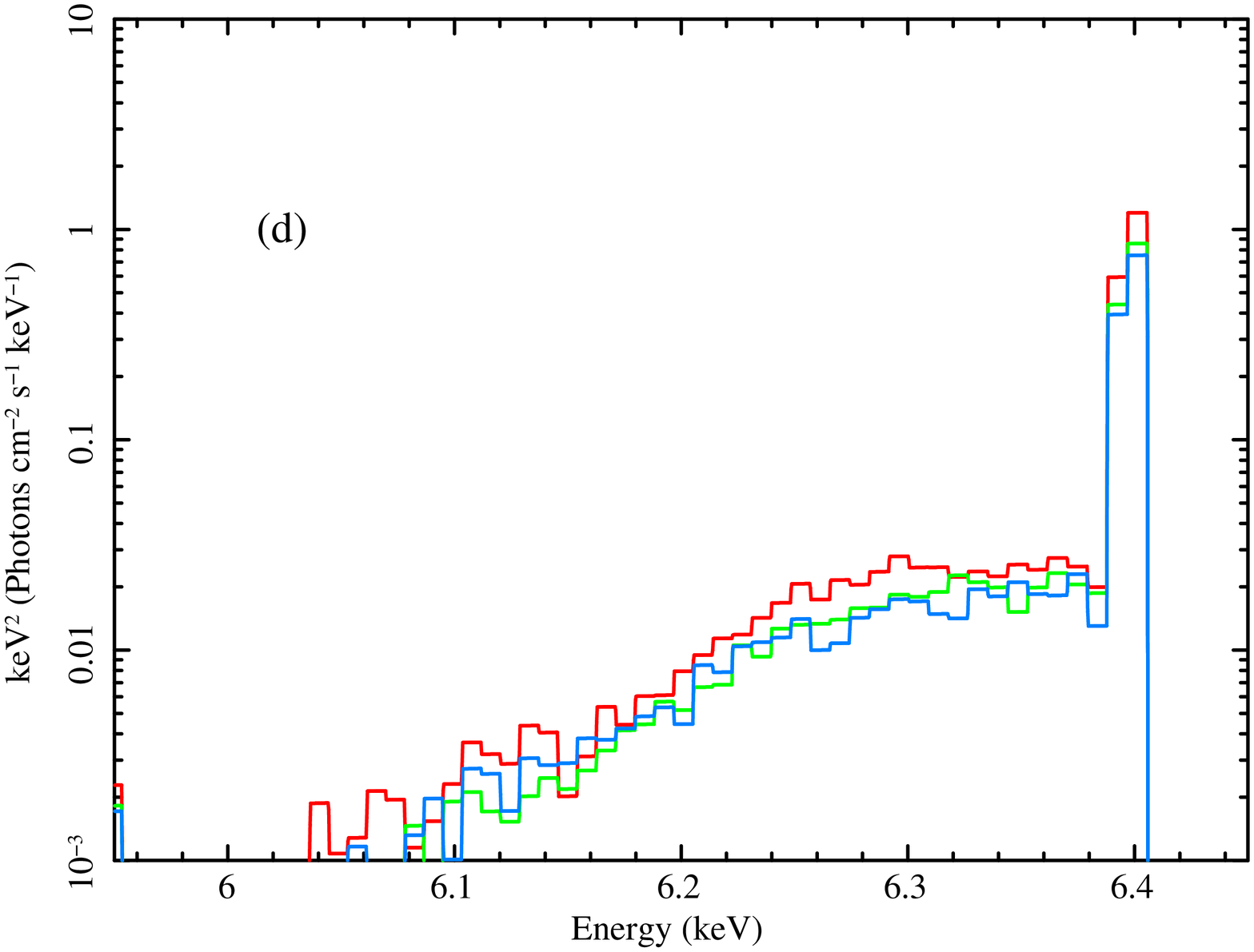}
\caption{Dependence on the number of clumps along the equatorial plane ($N_{\mathrm{clump}}^{\mathrm{Equ}}$) of (a) broadband X-ray spectrum for $\log N_{\mathrm{H}}/$cm$^{-2} = 24.0$, (b) Fe K$\alpha$ line profile for $\log N_{\mathrm{H}}/$cm$^{-2} = 24.0$, (c) broadband X-ray spectrum for $\log N_{\mathrm{H}}/$cm$^{-2} = 25.0$, and (d) Fe K$\alpha$ line profile for $\log N_{\mathrm{H}}/$cm$^{-2} = 25.0$. We adopt the following default parameters: $\sigma = 40.0$ degree, $i = 60.0$ degree, $\Gamma = 2.0$, and $E_{\mathrm{cut}} = 100$ keV. Red line: $N_{\mathrm{clump}}^{\mathrm{Equ}} = 5.0$, Green line: $N_{\mathrm{clump}}^{\mathrm{Equ}} = 10.0$. Blue line: $N_{\mathrm{clump}}^{\mathrm{Equ}} = 15.0$}
\end{figure*}
\begin{figure*}
\plottwo{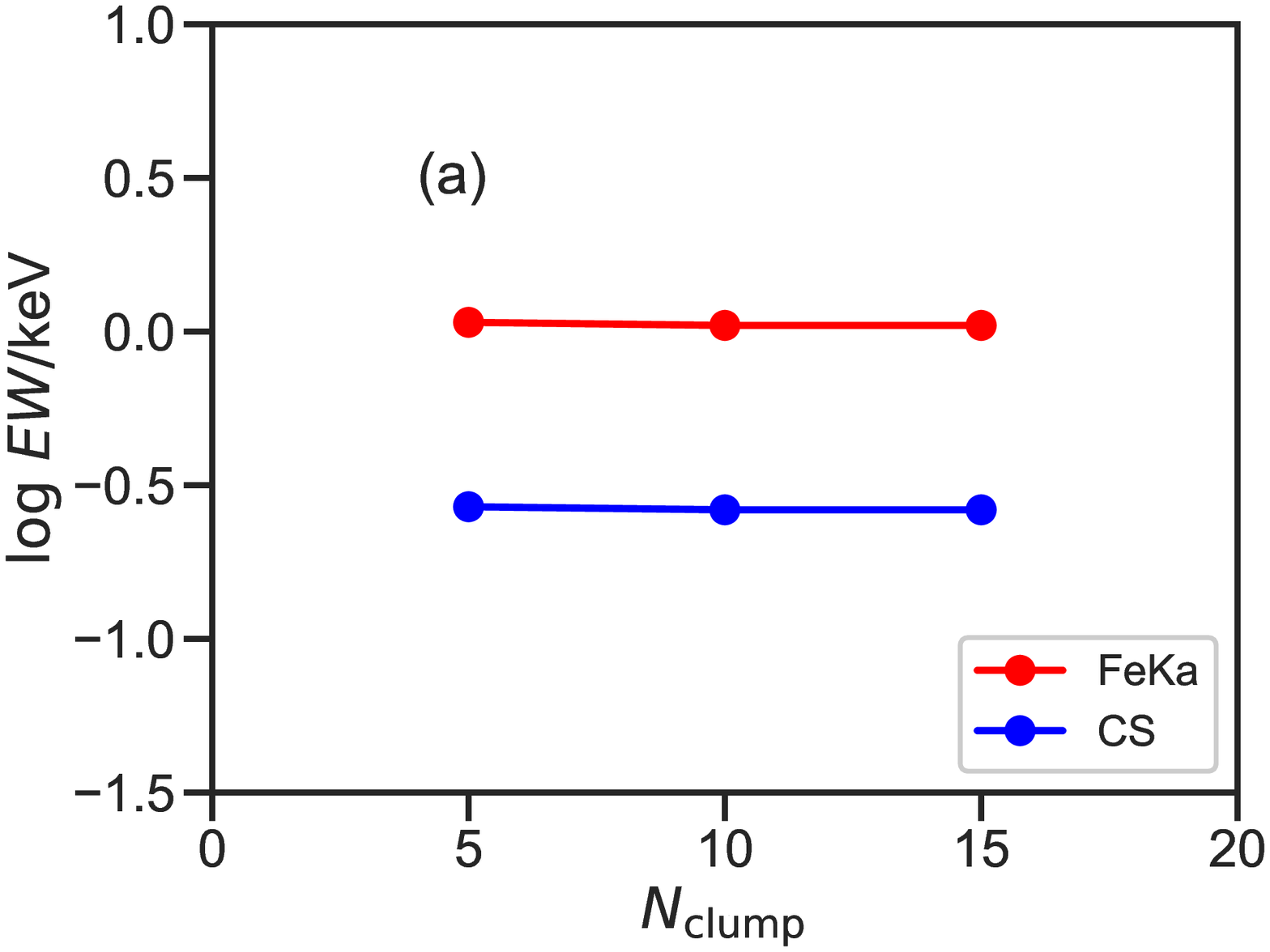}{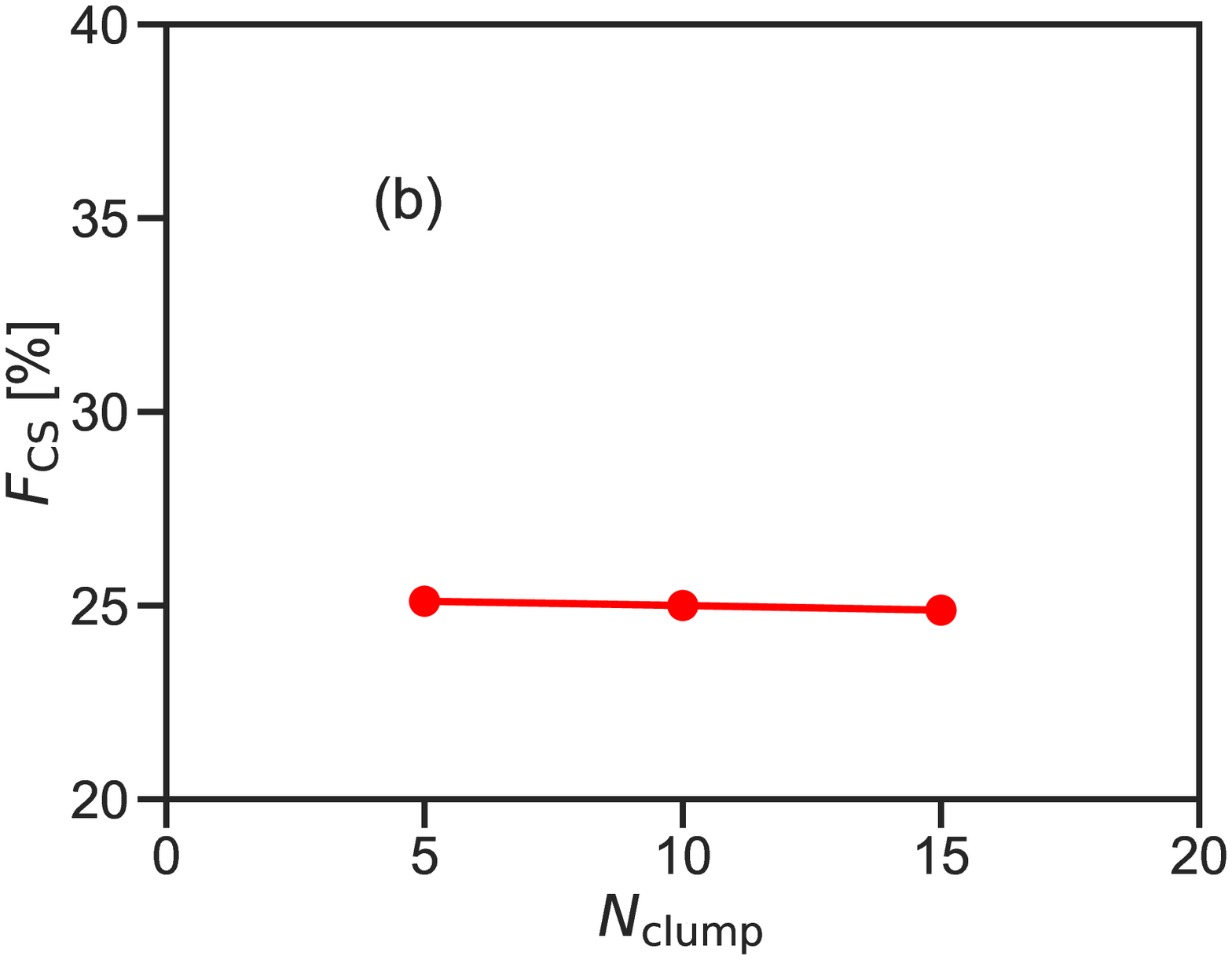}
\plottwo{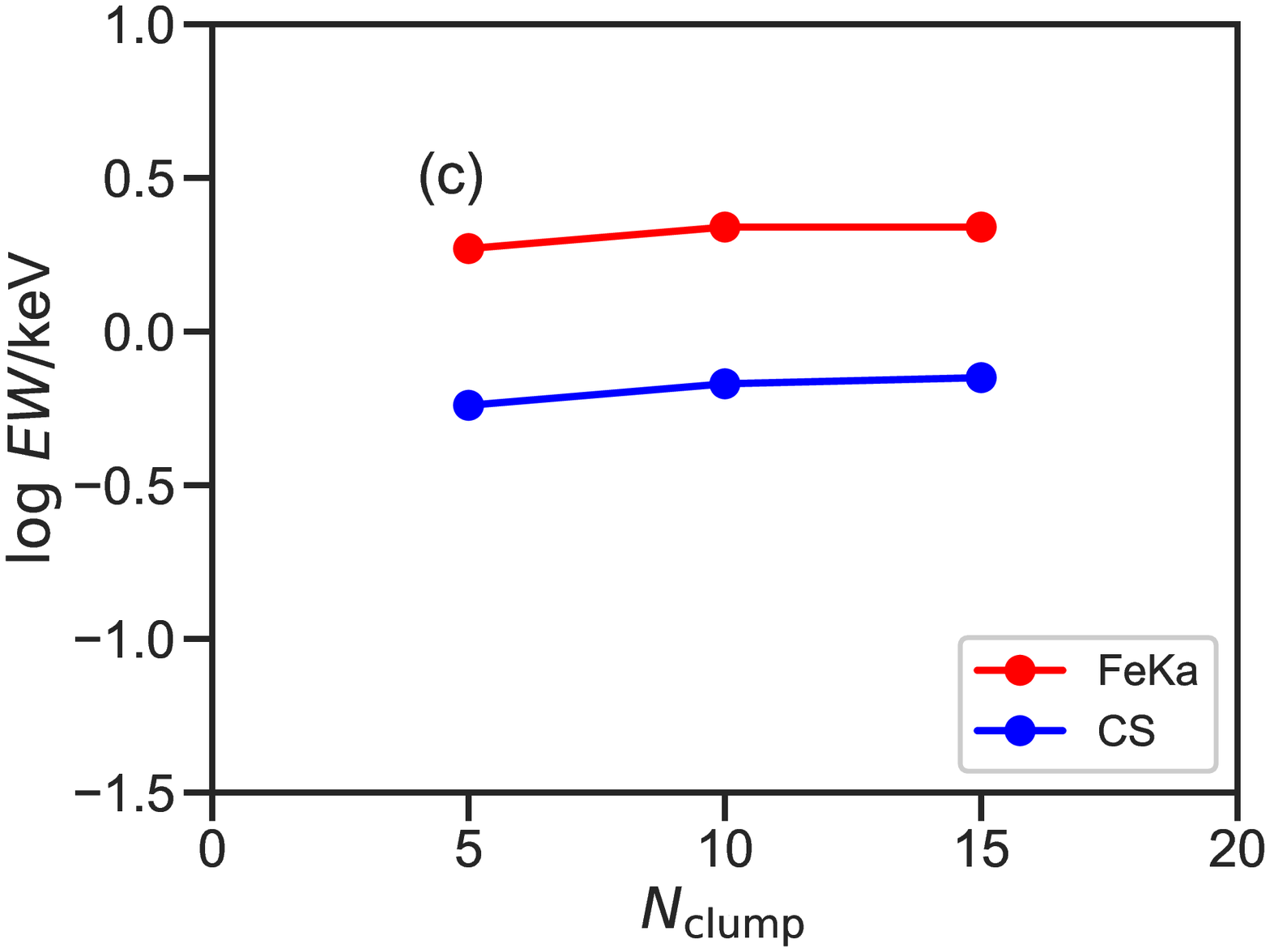}{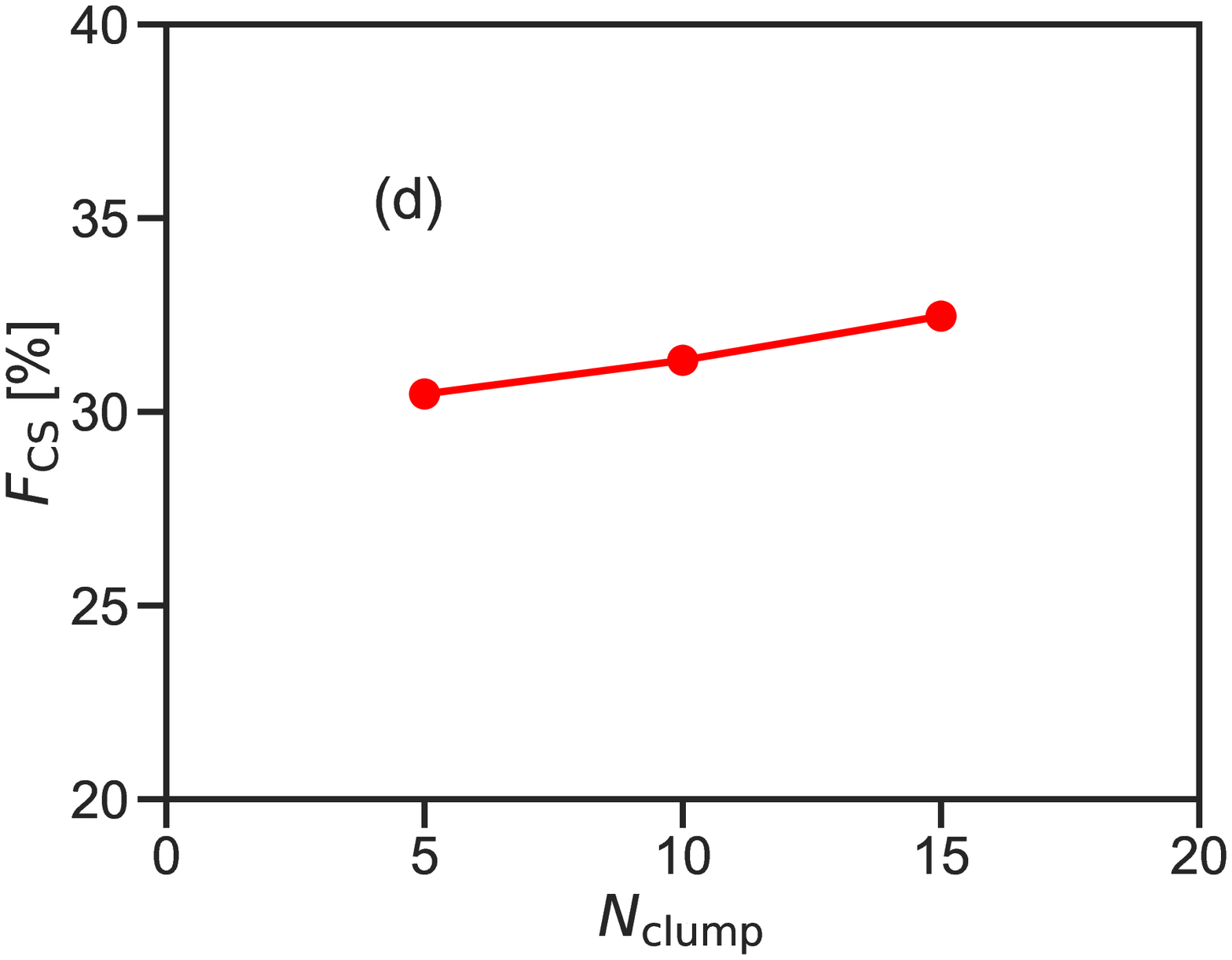}
\caption{Dependence of equivalent widths and Compton shoulder fraction on the number of clumps along the equatorial plane ($N_{\mathrm{clump}}^{\mathrm{Equ}}$): (a)-(b) for $\log N_{\mathrm{H}}/$cm$^{-2} = 24.0$. (c)-(d) for $\log N_{\mathrm{H}}/$cm$^{-2} = 25.0$. We set the following default parameters: $\sigma = 40.0$ degree, $i = 60.0$ degree, $\Gamma = 2.0$, and $E_{\mathrm{cut}} = 100$ keV. Left: Red line: equivalent width of Fe K$\alpha$. Blue line: equivalent width of Compton shoulder. Right: Red line: Compton shoulder fraction.}
\end{figure*}
\begin{figure*}
\plottwo{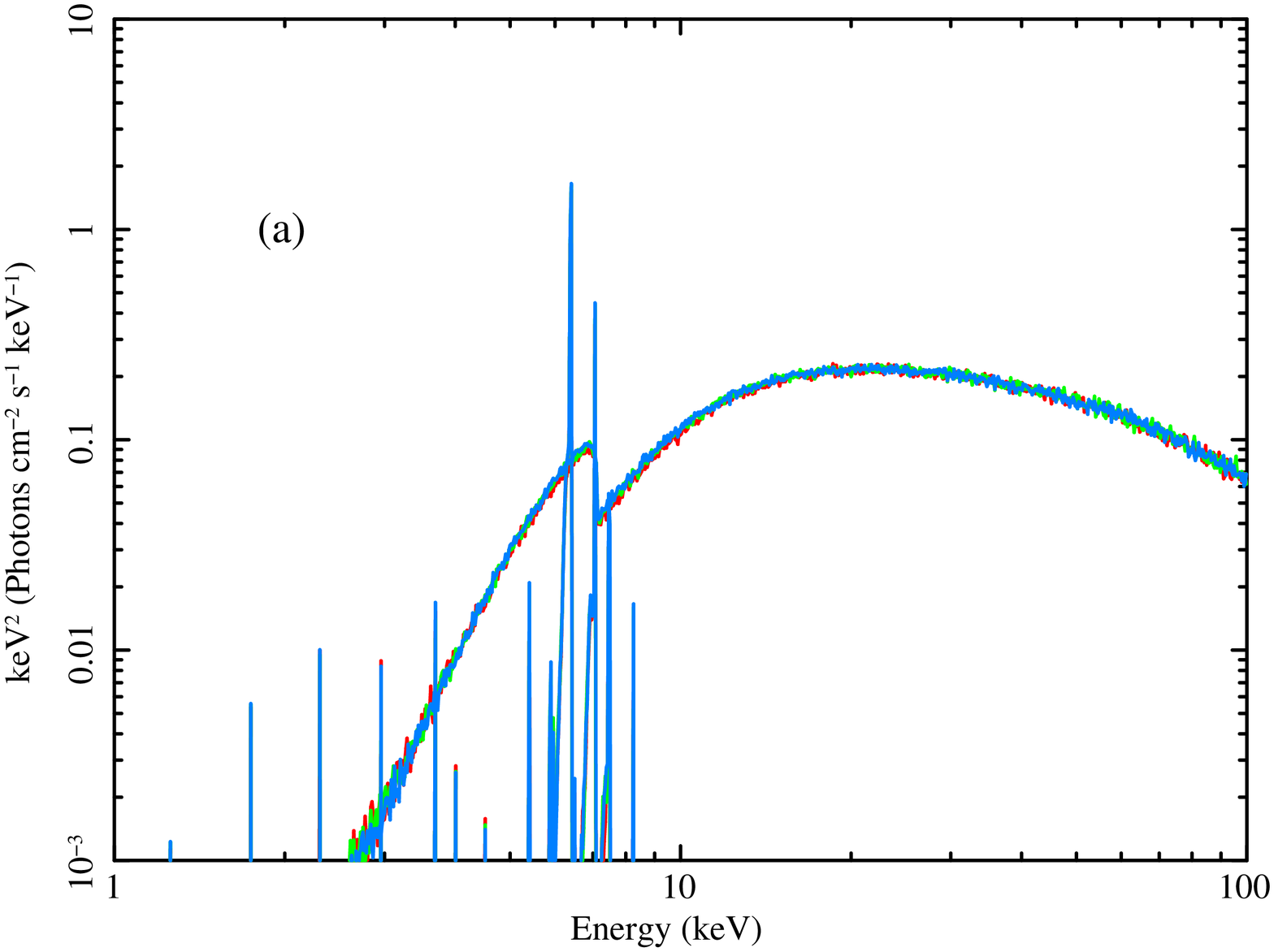}{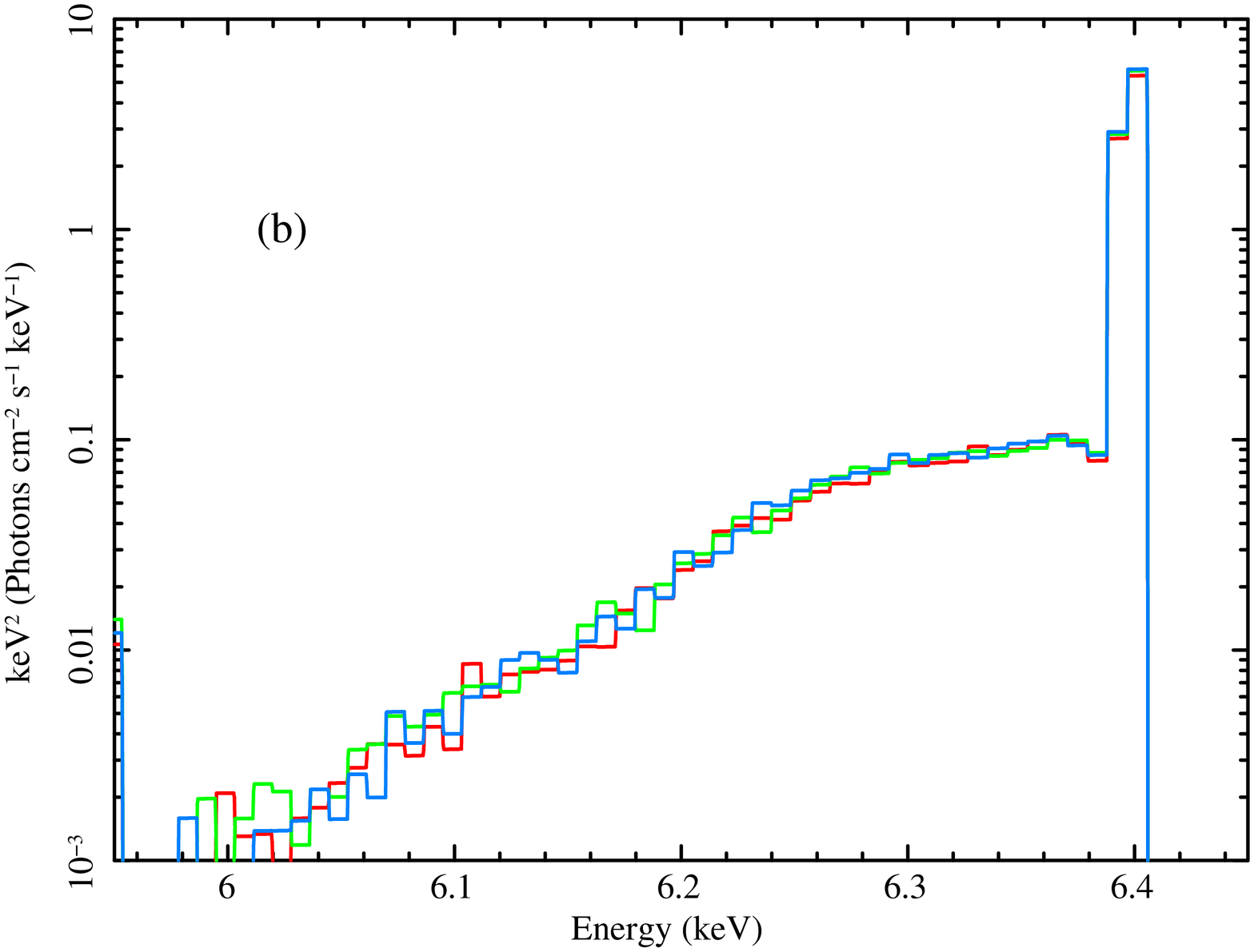}
\plottwo{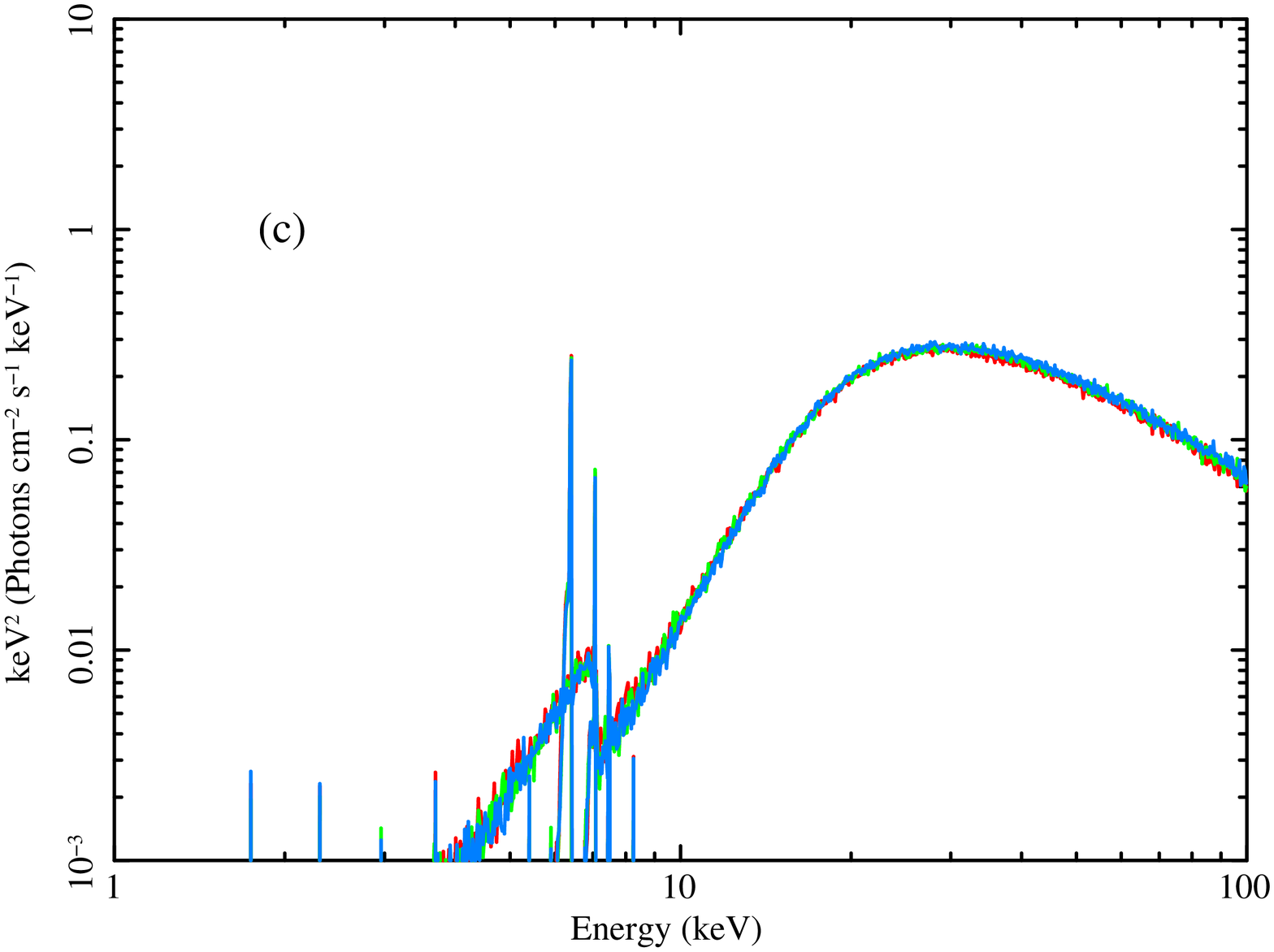}{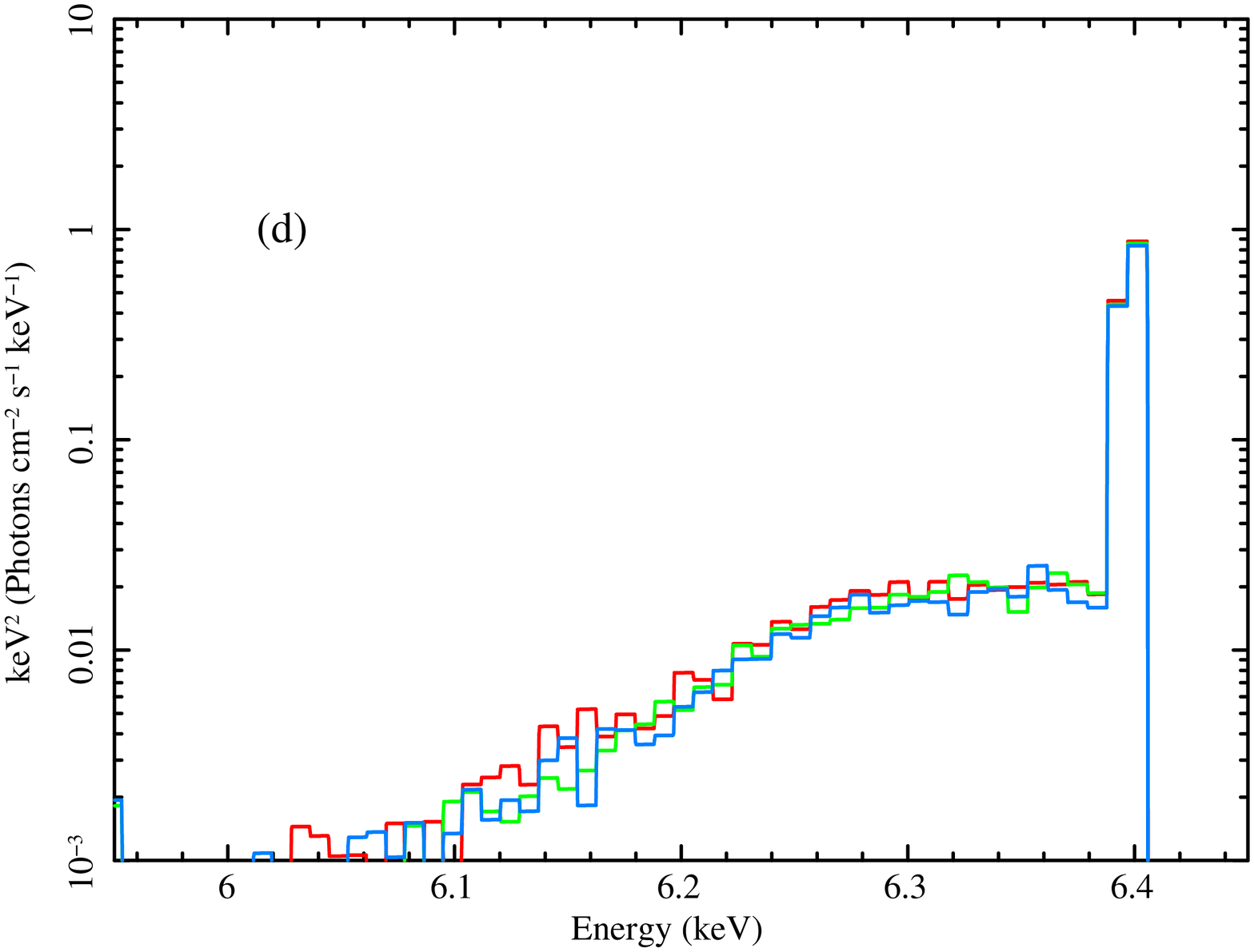}
\caption{Dependence on the radial density profile ($q$) of (a) broadband X-ray spectrum for $\log N_{\mathrm{H}}/$cm$^{-2} = 24.0$, (b) Fe K$\alpha$ line profile for $\log N_{\mathrm{H}}/$cm$^{-2} = 24.0$, (c) broadband X-ray spectrum for $\log N_{\mathrm{H}}/$cm$^{-2} = 25.0$, and (d) Fe K$\alpha$ line profile for $\log N_{\mathrm{H}}/$cm$^{-2} = 25.0$. We adopt the following default parameters: $\sigma = 40.0$ degree, $i = 60.0$ degree, $\Gamma = 2.0$, and $E_{\mathrm{cut}} = 100$ keV. Red line: $q = 0.0$. Green line: $q = 0.5$. Blue line: $q = 1.0$.}
\end{figure*}
\begin{figure*}
\plottwo{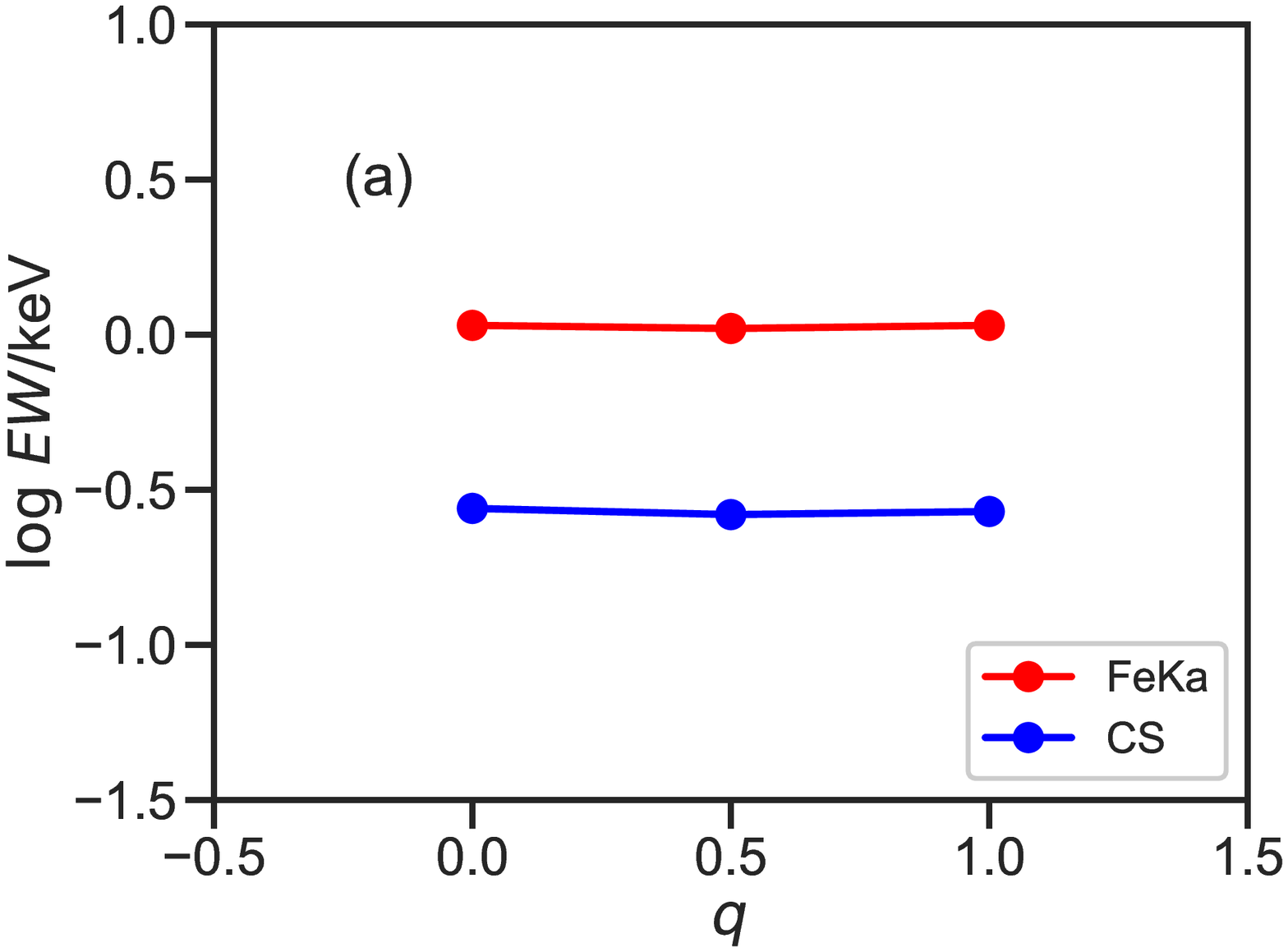}{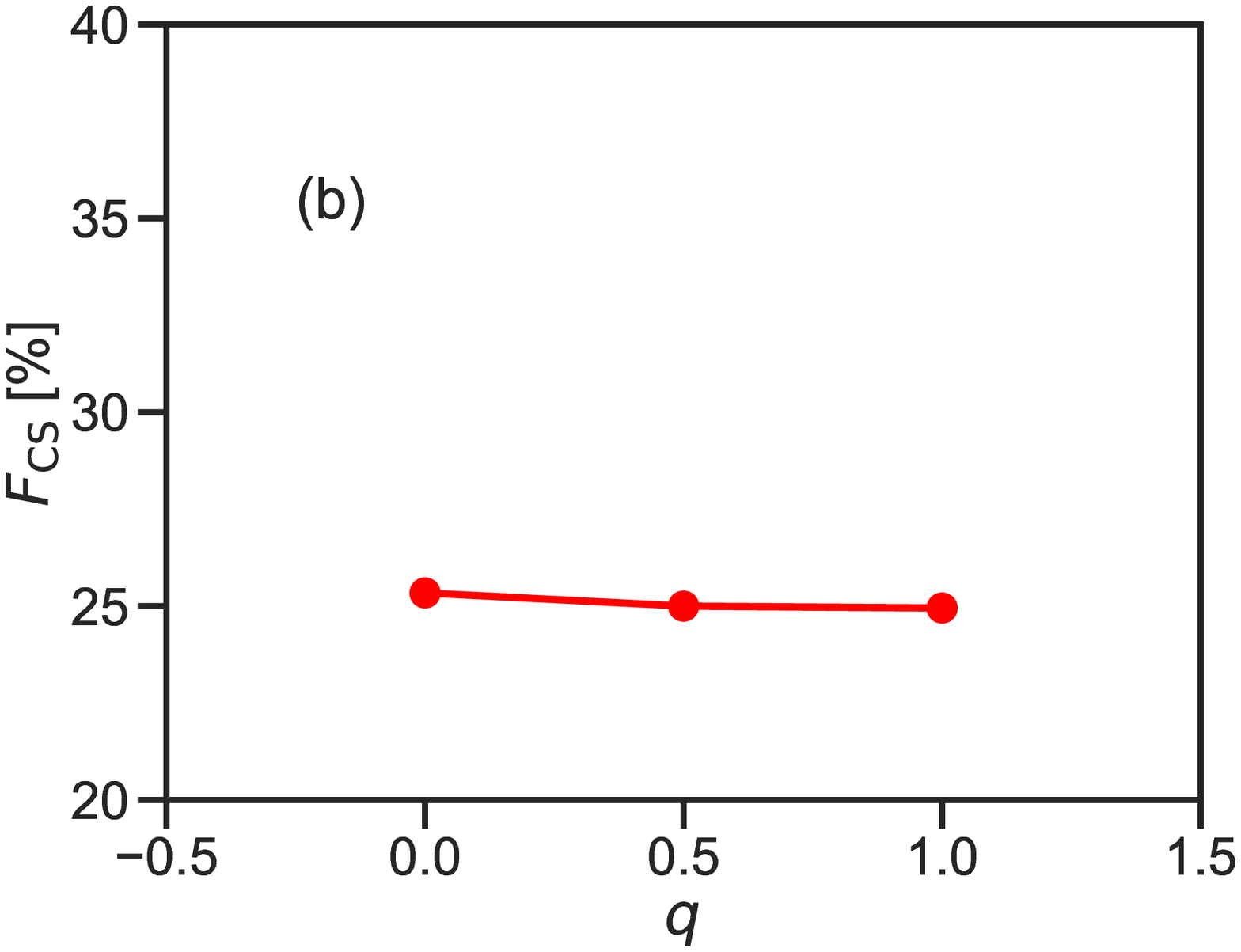}
\plottwo{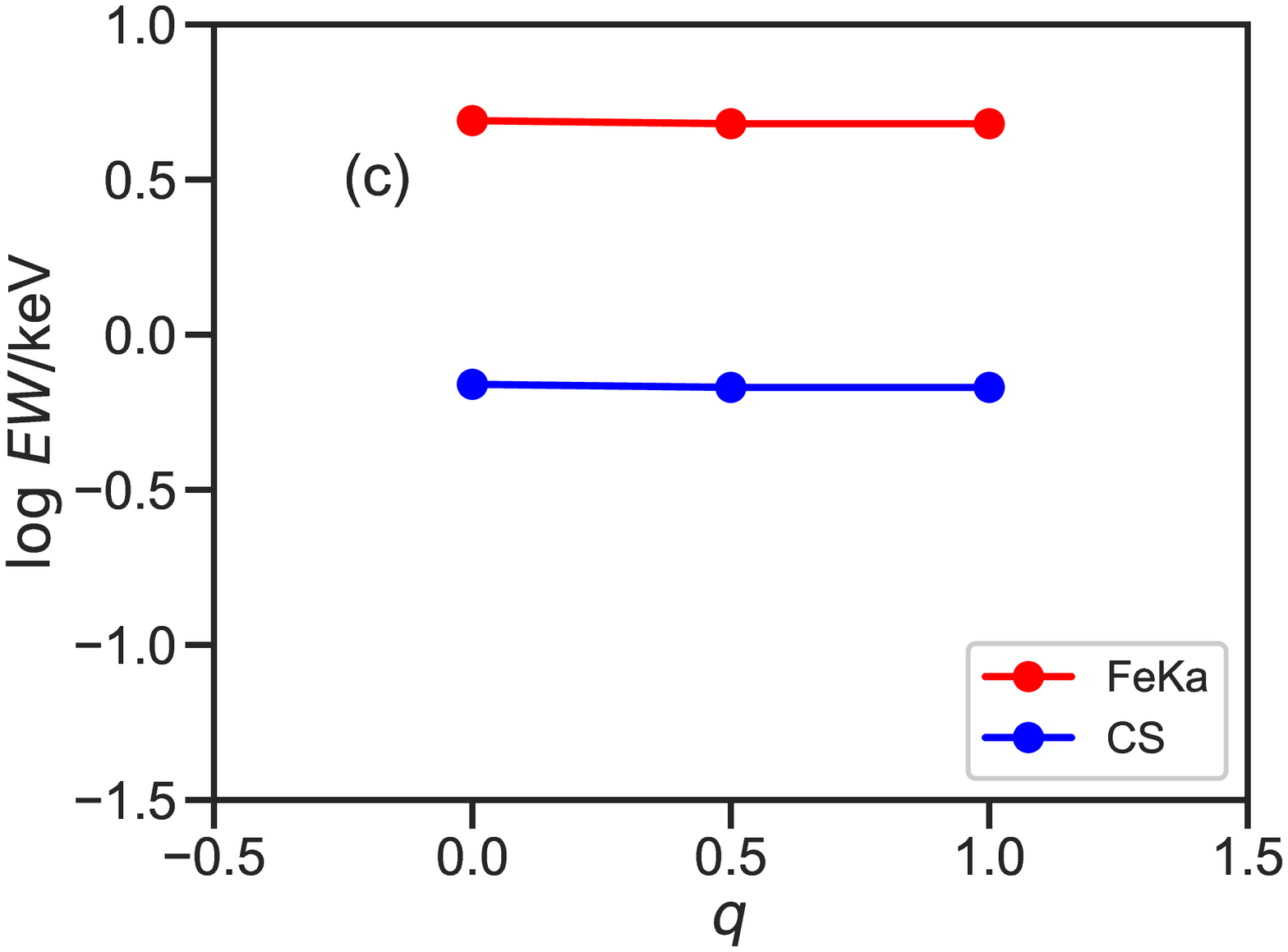}{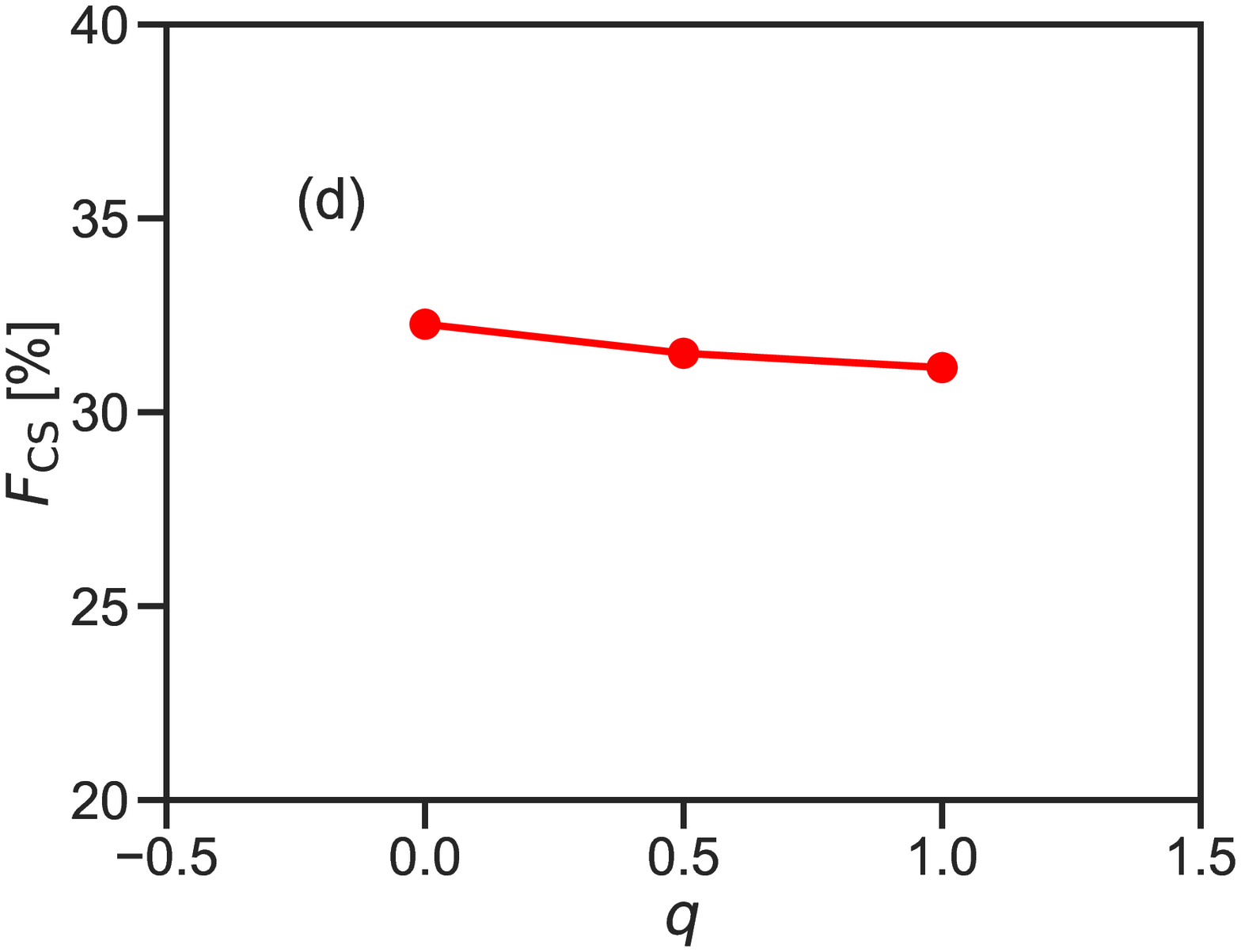}
\caption{Dependence of equivalent widths and Compton shoulder fraction on the radial density profile ($q$). (a)-(b) for $\log N_{\mathrm{H}}/$cm$^{-2} = 24.0$. (c)-(d) for $\log N_{\mathrm{H}}/$cm$^{-2} = 25.0$. We set the following default parameters: $\sigma = 40.0$ degree, $i = 60.0$ degree, $\Gamma = 2.0$, and $E_{\mathrm{cut}} = 100$ keV. Left: Red line: equivalent width of Fe K$\alpha$. Blue line: equivalent width of Compton shoulder. Right: Red line: Compton shoulder fraction.}
\end{figure*}

\appendix
\section{Dependence on the number of clumps along the equatorial plane}
We investigate the dependence of the spectrum on the number of clumps along the equatorial plane ($N_{\mathrm{clump}}^{\mathrm{Equ}}$). Here we adopt the following default parameters: $\sigma = 40.0$ degree, $i = 60.0$ degree, $\Gamma = 2.0$, and $E_{\mathrm{cut}} = 100$ keV. Figure 8 compares the broadband X-ray spectrum and Fe K$\alpha$ line profile among different $N_{\mathrm{clump}}^{\mathrm{Equ}}$ values (5, 10, and 15) for $\log N_{\mathrm{H}}/$cm$^{-2} =$ 24.0 and 25.0. As noticed, the spectrum little depends on $N_{\mathrm{clump}}^{\mathrm{Equ}}$ for $\log N_{\mathrm{H}}/$cm$^{-2} = 24.0$. By contrast, for $\log N_{\mathrm{H}}/$cm$^{-2} = 25.0$, the continuum flux below 20 keV and Fe K$\alpha$ line flux  become higher with $N_{\mathrm{clump}}^{\mathrm{Equ}} = 5.0$ than with  $N_{\mathrm{clump}}^{\mathrm{Equ}} = 10.0$ or 15.0; the Fe K$\alpha$ line flux with $N_{\mathrm{clump}}^{\mathrm{Equ}} = 5.0$ is 25\% higher than that with $N_{\mathrm{clump}}^{\mathrm{Equ}} = 10.0$. The same trend is also reported in \cite{Liu14}. This is because the smaller number of clumps in the line-of-sight works to reduce the total probability of absorption for soft X-rays when each clump is already optically thick. Figure 9 plots the dependence of equivalent widths and Compton shoulder fraction on $N_{\mathrm{clump}}^{\mathrm{Equ}}$ for $\log N_{\mathrm{H}}/$cm$^{-2} = 24.0$ and $\log N_{\mathrm{H}}/$cm$^{-2} = 25.0$. We find that they show little dependence on $N_{\mathrm{clump}}^{\mathrm{Equ}}$ both for $\log N_{\mathrm{H}}/$cm$^{-2} = 24.0$ and 25.0. This is because the changes of the continuum flux and  Fe K$\alpha$ line flux are almost cancelled out each other.

\section{Dependence on the index of the radial density profile}
We also investigate the dependence of the spectrum on the index of the radial density profile ($q$). We adopt the following default parameters: $\sigma = 40.0$ degree, $i = 60.0$ degree, $\Gamma = 2.0$, and $E_{\mathrm{cut}} = 100$ keV). Figure 10 compares the broadband X-ray spectrum and Fe K$\alpha$ line profile among different among different $q$ values (0.0, 0.5, 1.0) for $\log N_{\mathrm{H}}/$cm$^{-2} =$ 24.0 and 25.0. Figure 11 plots the dependence of equivalent widths and Compton shoulder fraction on $q$ for $\log N_{\mathrm{H}}/$cm$^{-2} =$ 24.0 and 25.0. We find that they show little dependence on $q$ within a range of 0.0--1.0.

\bibliographystyle{aasjournal}
\bibliography{tanimoto}
\end{document}